\documentclass[aps,prd,twocolumn,showpacs,eqsecnum,nofootinbib,amsmath,amssymb,amsfonts,english]{revtex4}

\usepackage{babel}
\usepackage{graphicx}

\newcommand{\px}{{\mathrm{px}}}
\newcommand{\pxI}{{\mathrm{pxI}}}
\newcommand{\pxII}{{\mathrm{pxII}}}
\newcommand{\fr}{{\mathrm{free}}}
\newcommand{\mx}{{\mathrm{max}}}
\newcommand{\mxI}{{\mathrm{maxI}}}
\newcommand{\mxII}{{\mathrm{maxII}}}
\newcommand{\inc}{{\mathrm{in}}}
\newcommand{\out}{{\mathrm{out}}}

\newcommand{\ph}{\varphi}

\newcommand{\figdir}{}

\begin{document}

\title{Billiard in the space with a time machine}

\author{Jind\v{r}ich Dolansk\'{y}}
\email{dolanskyy@gmail.com}
\author{Pavel Krtou\v{s}}
\email{Pavel.Krtous@mff.cuni.cz}

\affiliation{
  Institute of Theoretical Physics,
  Faculty of Mathematics and Physics, Charles University in Prague,\\
  V Hole\v{s}ovi\v{c}k\'{a}ch 2, 180 00 Prague 8, Czech Republic
  }

\date{October 11, 2010} 

\begin{abstract}
We study a system of an elastic ball moving in the non-relativistic spacetime with a nontrivial causal structure produced by a wormhole-based time machine. For such a system it is possible to formulate a simple model of the so-called ``grandfather paradox'': for certain ``paradoxical'' initial conditions the standard straight trajectory of the ball would self-collide inconsistently. We analyze globally consistent solutions of local equations of motion, namely, we find all trajectories with one self-collision. It is demonstrated that all standard initial conditions have a consistent evolution, including those ``paradoxical'' ones, for which the inconsistent collision-free trajectory is superseded by a special consistent self-colliding trajectory. Moreover, it is shown that for a wide class of initial conditions more than one globally consistent evolution exist. The nontrivial causal structure thus breaks the uniqueness of the classical theory even for locally deterministic physical laws.
\end{abstract}



\maketitle


\section{Introduction}
\label{sc:intro}

Time travel is a phenomenon which has been attracting interest both in fiction and general discussions for a long time. However, only after a formulation of the theory of relativity such considerations could be investigated on a more scientific and solid basis. Already the special relativity shows that different observers experience different times and one of them can ``travel'' to the future of others by means of his relative motion. Thanks to the general theory of relativity a possibility opens that an observer could travel even to his own past -- his worldline could pass through a geometrically or topologically nontrivial area to a region where the worldline originally started \cite{tipler,godel}. Worldlines which even cross themselves are called \emph{closed timelike curves (CTCs)} and it is customary to say that spacetimes with CTCs contain \emph{time machines} \cite{wormhole,cosmic-string}.

Spacetimes with time machines are causally nontrivial -- in such spacetimes you can send a signal to your own past or even try to influence the past -- which immediately opens a question of consistency of standard physical laws as we know them. On a formal level it is the question of the existence of solutions of physical equations of motion and the question whether the initial value problem is well possessed. On a less formal level these problems can be phrased as the well-known ``grandfather paradox'', suggested, e.g., in \cite{Thorne:book,Gott:book,lorentzian-whs}: in spacetimes with time machines one has to face a logical riddle what happens if one travels to his own past and kills his grandfather. Consequently one would never be born and therefore one could not travel to the past.

This is a clearly inconsistent situation which suggests that spacetimes with CTCs are pathological and they should be excluded from a serious scientific consideration. However, a system containing live beings is too complicated by too many unknown physical laws and therefore one cannot be sure that the inconsistency of the ``grandfather paradox'' is really inescapable. Therefore, people tried to formulate analogous situations for much simpler systems which could be studied exactly \cite{novikov,friedman,cauchy,tm.ec}.

\begin{figure}[b]\begin{center}
\includegraphics[width=3.3in]{\figdir 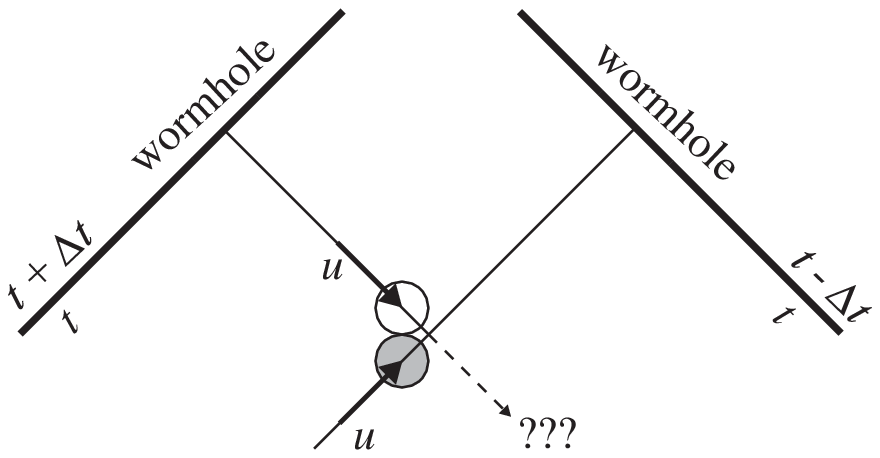}\caption{\label{fig:dangerous-px}\small%
\textbf{The ball self-colliding inconsistently.} The ball (in gray) comes from a distant region and enters the wormhole without any self-collision. It leaves the time machine (now in white) in such a way that it inconsistently hits itself. Thus, this situation represents an inconsistent evolution which is the direct analogue of the grandfather paradox.}%
\end{center}\end{figure}

As an example, in \cite{mensky}, and especially in \cite{billiard}, the system of the billiard balls in spacetime with wormholes has been studied. This system allows a straightforward reformulation of the grandfather paradox: the ball could be sent through the time machine in such a way that it hits itself and thus prevents inconsistently its entry to the time machine, cf.\ Fig.~\ref{fig:dangerous-px}. It seems that such ``paradoxical'' initial conditions do not lead to a consistent evolution of the system.

Maybe surprisingly, the extensive studies of systems with CTCs during the 1980s and the 1990s showed, that for a simple physical system pathology of spacetimes is not so severe and the equations of motion can be consistently solved.

Let us formulate this point more precisely. We consider a spacetime containing a time machine and we want to study a system with well known local physical laws (e.g., a particle or electromagnetic field). We do not change these local laws, i.e., we require that they hold locally in any small spacetime domain. However, in addition to the local laws, we also require the so-called \emph{principle of self-consistency} \cite{novikov,self-consistency,cauchy}. Namely, a \emph{globally consistent} solution of local laws must exist. It means that we allow the system to propagate to its own past, however, it must be done in a \emph{consistent} way with the original evolution in the past. The past cannot be changed since it has been already changed.

The key question of studies of time machines is whether such globally consistent evolutions exist for given local laws and whether these global evolutions are sufficiently generic. More accurately, we would like to show that there exist consistent solutions for all, or, at least, for almost all standard initial data. Otherwise, if the local laws have no globally consistent solution, the spacetime would be clearly pathological and we could rule it out from our consideration. Similarly, the pathology would be serious if the local physical laws had only few globally consistent solutions.

As we have said, the studies of different systems show that spacetimes with CTCs are not necessarily causally pathological. Let us mention the results for a system of interacting particles \cite{billiard,novikov,mensky} or the scalar field theory \cite{friedman,cauchy} where it was shown that standard local laws have generic globally consistent solutions even in the presence of CTCs. Another surprising result of such studies is that the existence of time machines does not usually restrict a number of consistent solutions, but on the contrary, it leads to a possibility of more than one globally consistent solutions for given initial values. In spacetimes with time machines we thus usually lose the uniqueness of the evolution \cite{billiard}.

Especially for the apparently paradoxical initial conditions from the ``grandfather paradox''-like situation it was shown \cite{billiard} that a consistent evolution exists, although it can be rather nonintuitive. On less formal level of the human version of the paradox it could be rephrased as a conjecture of a hidden law which always prevents the grandson from killing his grandfather \cite{kras.tm}.

In the present work we want to study the system of a ball moving in a non-relativistic space with CTCs which is very close to that of \cite{billiard}. However, we have chosen a different time-machine configuration which significantly simplifies the analysis. We will be able to solve the equations of motion explicitly and we will confirm the behavior described above: ``paradoxical'' initial conditions are not really paradoxical, and the evolution is not, in general, unique. The contribution of our analysis is that it can be done very explicitly, since our model is sufficiently simple; at the same time it is sufficiently non-trivial to reproduce the most important features of systems with time machines.

Our work proceeds as follows. In the next section we describe the system of a ball moving in the space with wormhole-based time machine. We derive the equations characterizing trajectories. In Sec.~\ref{sc:numbsol} we discuss the character and number of solutions for various initial conditions and also the resolution of the ``grandfather paradox''. Sec.~\ref{sc:ppart} describes geometry of the trajectories and the paper is summarized in the conclusion.

\section{Description of the system}
\label{sc:system}

\subsection*{Wormhole time machine}

\begin{figure}\begin{center}
\includegraphics[width=3.3in]{\figdir 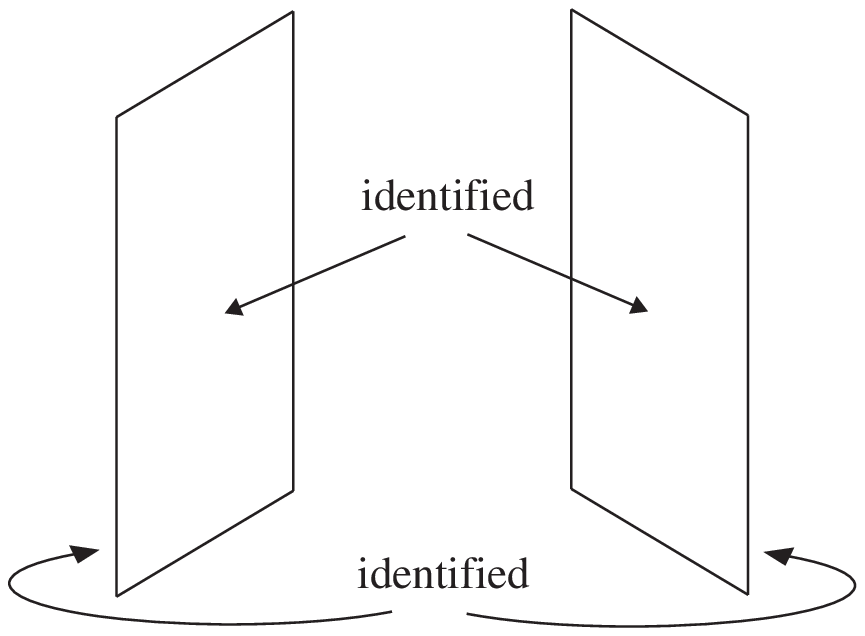}
\caption{\label{fig:wh-s}\small%
\textbf{Spatial representation of two simple wormholes.}
A wormhole obtained by gluing two planar sections cut from otherwise Euclidian space. Planar wormhole mouths could be obtained, e.g., by squeezing the mouths of the spherical wormhole into very thin planes). The external curvature is vanishing with the exception of the boundary of the planar sections and the geometry through the wormhole is thus flat.}\vspace{-3ex}
\end{center}\end{figure}

The simplest and most natural way how to construct a spacetime with CTCs is to use a wormhole \cite{lorentzian-whs,tm.ec,billiard}. The wormhole can be viewed as a shortcut between two location of the spacetime. It forms CTCs if one of the mouths of the wormhole lies in the past of the other mouth.

In the nonrelativistic setting we can consider spatial wormholes connecting two places in space with an additional time-shift. The speed of light is infinite and it determines a unique notion of simultaneity and thus it allows us to define a global time -- at least, before introducing the time difference. Introducing the time-shift means that traveling through the wormhole not only sends the observer to a different place in space but also to a different time.

A simple spatial wormhole in otherwise Euclidian space can be obtained by the cutting and gluing method. For example, we can cut out two spheres and glue their surfaces together, cf.\ \cite{wormhole}, or \cite{billiard}. We thus obtain a topologically and geometrically nontrivial space---it is not a simply connected space and the geometry on the glued surface is not flat.

In this work we consider even a simpler situation of the wormhole with planar mouths instead of spherical ones. Namely, we cut out from the space two planar sections which we identify as in Fig.~\ref{fig:wh-s}. Since we use flat planar sections, their identification is geometrically trivial. The whole curvature of the mouths is squeezed to the borders of the planar sections which can be understood as a kind of solid frames on which the traversable parts of the wormhole are spanned. To avoid a discussion of the wormhole boundary we assume that the planar sections are much larger than the scales of our experiments. As idealization, we consider the mouths of our wormhole to be two half-planes which form an angle ${\gamma}$ with a common boundary line, called the axis.

If we identify these half-planes (first, say, at the same moment of time), the space between them becomes a locally Euclidean space with a conical singularity localized on the axis. Indeed, if we restrict ourselves to the two dimensional picture and ignore the direction parallel to the axis, our space forms a cone with the angle ${\gamma}$ around the vertex.

Of course, this is a over-idealized situation. We should keep in mind that the mouths of the wormhole are large but finite, so somewhere very far from the axis the conical part of the space ends and goes over to the full Euclidean space. But in most of our discussions we restrict ourselves only to the part of the space between the mouths of the wormhole. We thus effectively work in the conical space with angle ${\gamma}$ around the axis.

Let us stress that in our construction the mouths of the wormhole are special and privileged. However, after enlarging them to the semi-infinite size and restricting ourselves only to the conical space between mouths, we can no longer localize the position of the mouths by local experiments. The geometry through the mouths is locally Euclidean as anywhere else. We thus obtained a space which is axially symmetric with respect to the rotation around the axis (it has also translation symmetry along the axis and it is static). The position of the wormhole can be identified only on scales larger than the wormhole, from the surrounding globally Euclidean space.

Finally, we assume that the angle ${\gamma\in(\frac\pi2,\pi)}$. Only for ${\gamma}$ smaller than ${\pi}$, we obtain interesting situations when the straight trajectory of a free particle intersects itself. For ${\gamma>\pi/2}$ the straight trajectory intersects itself only once. The second condition is not a crucial assumption, it just simplifies the discussion.

Since we want to study a space with a time machine, we have to identify the mouths of the wormhole with a time-shift ${\Delta t}$.
Of course, it destroys the standard causal structure of the non-relativistic spacetime (a clear distinction between future and past). Despite this, we will keep using and referring to the original notion of the simultaneity and to the global time of the surrounding space, which is, certainly, reasonable for small wormholes, but we admit, it can be slightly confusing and less founded in the idealized case considered below.

Indeed, the constructed spacetime is still locally Euclidian (of course, except the axis), but endowed with a strange causal structure. Hypersurfaces of simultaneity (visualized in the standard non-relativistic spacetime diagram as horizontal planes) propagate through the wormhole and form ``helical'' surfaces winding around the axis. This indicates that the spacetime contains CTCs.
The particle moving towards the wormhole reads that the external time $t$ increases continuously until it enters the wormhole mouth. By crossing one of the mouths the external time $t$ either increases by $\Delta t$ into $t+\Delta t$, or decreases by $-\Delta t$ into $t-\Delta t$.

\begin{figure}\begin{center}
\includegraphics[width=3.3in]{\figdir 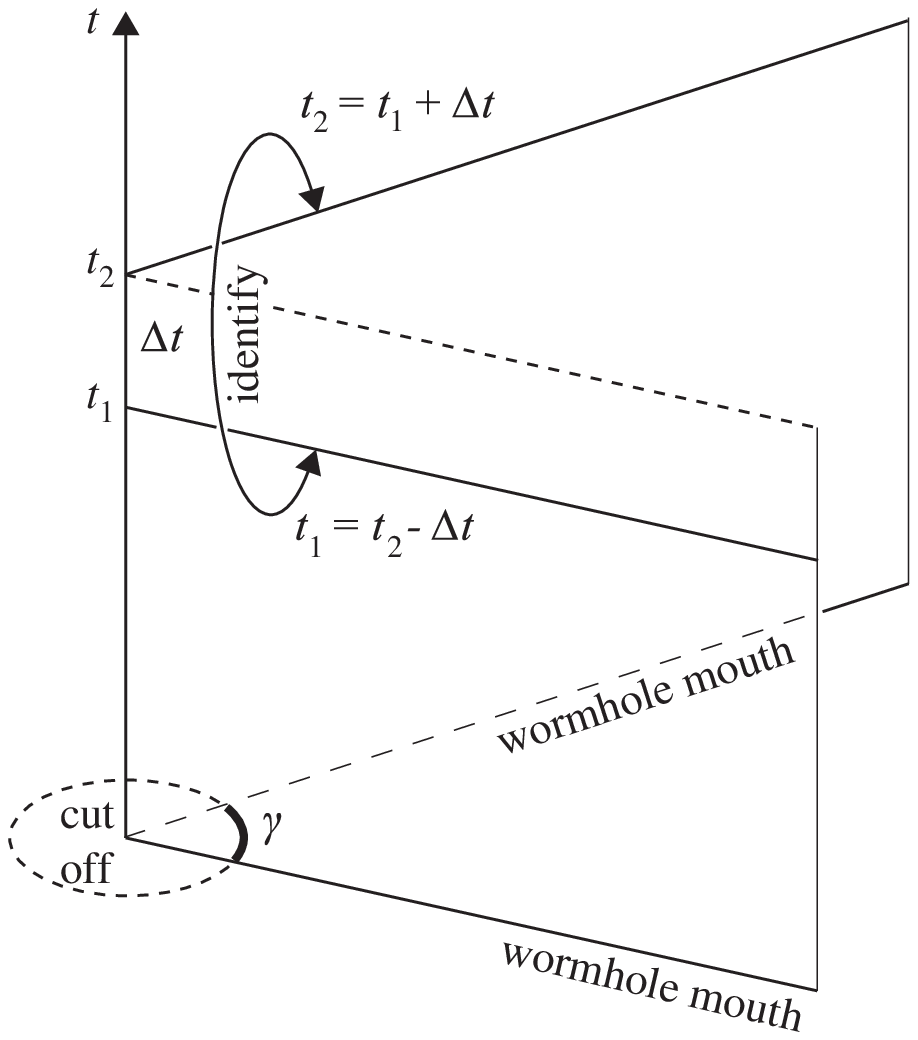}
\caption{\label{fig:wh-st}\small%
\textbf{A spacetime diagram of the conical time machine.}
The vertical direction is temporal, horizontal planes correspond to the hypersurface of simultaneity of the original spacetime (the third spatial direction is suppressed). Two half-planes on the boundary of the conical spacetime represent the history of the mouths of the wormhole and they are identified with a time-shift ${\Delta t}$.}%
\end{center}\end{figure}

This can be visualized in the spacetime diagram in Fig.~\ref{fig:wh-st}. Here, two spatial directions perpendicular to the axis are shown in horizontal directions, the direction parallel to axis is suppressed. The vertical direction corresponds to time. Semi-planar mouths of the wormhole at one moment are thus depicted as horizontal semi-lines, their time evolution as vertical half-planes. The identification of such two half-planes is not on the same vertical level, but with the vertical shift ${\Delta t}$. We assume that going through the wormhole in anticlockwise direction takes us time ${\Delta t>0}$ to the past, in clockwise direction to the future.

\subsection*{Equation of motion}

Now, we want to derive equations of motion for a ball moving in the non-relativistic spacetime which has just been described. We assume that the motion satisfies classical local laws of motion. Additionally, we restrict ourselves to motions perpendicular to the axis of the wormhole because the ball can self-interact only for such a motion.

Clearly, a free particle moves uniformly along a straight line. However, for ${\gamma\in(\pi/2,\pi)}$, a straight line in a conical space must intersect itself. If the trajectory of the particle crosses itself in different times, we speak about \emph{self-intersection}. If the particle intersects its trajectory exactly at the same time---which is allowed thanks to time-shift $-\Delta t$ gained in the wormhole---it hits itself and we speak about \emph{self-collision}. Our goal is to describe trajectories with exactly one consistent self-collision.
\begin{figure}\begin{center}
\includegraphics[width=3.3in]{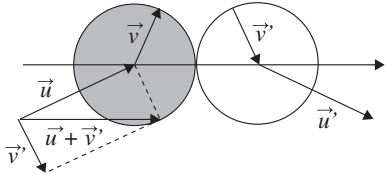}
\caption{\label{fig:impact}\small%
\textbf{Character of self-collision of the ball with itself.}
The incoming velocity ${\vec{u}}$ of the younger version of the ball (in grey) is complementary to mirror reflection of the outgoing velocity ${\vec{v}'}$ of the older version of the ball (white) and similarly for velocities ${\vec{u}'}$ and ${\vec{v}}$.}%
\end{center}\end{figure}

\begin{figure}\begin{center}
\includegraphics[width=3.3in]{\figdir 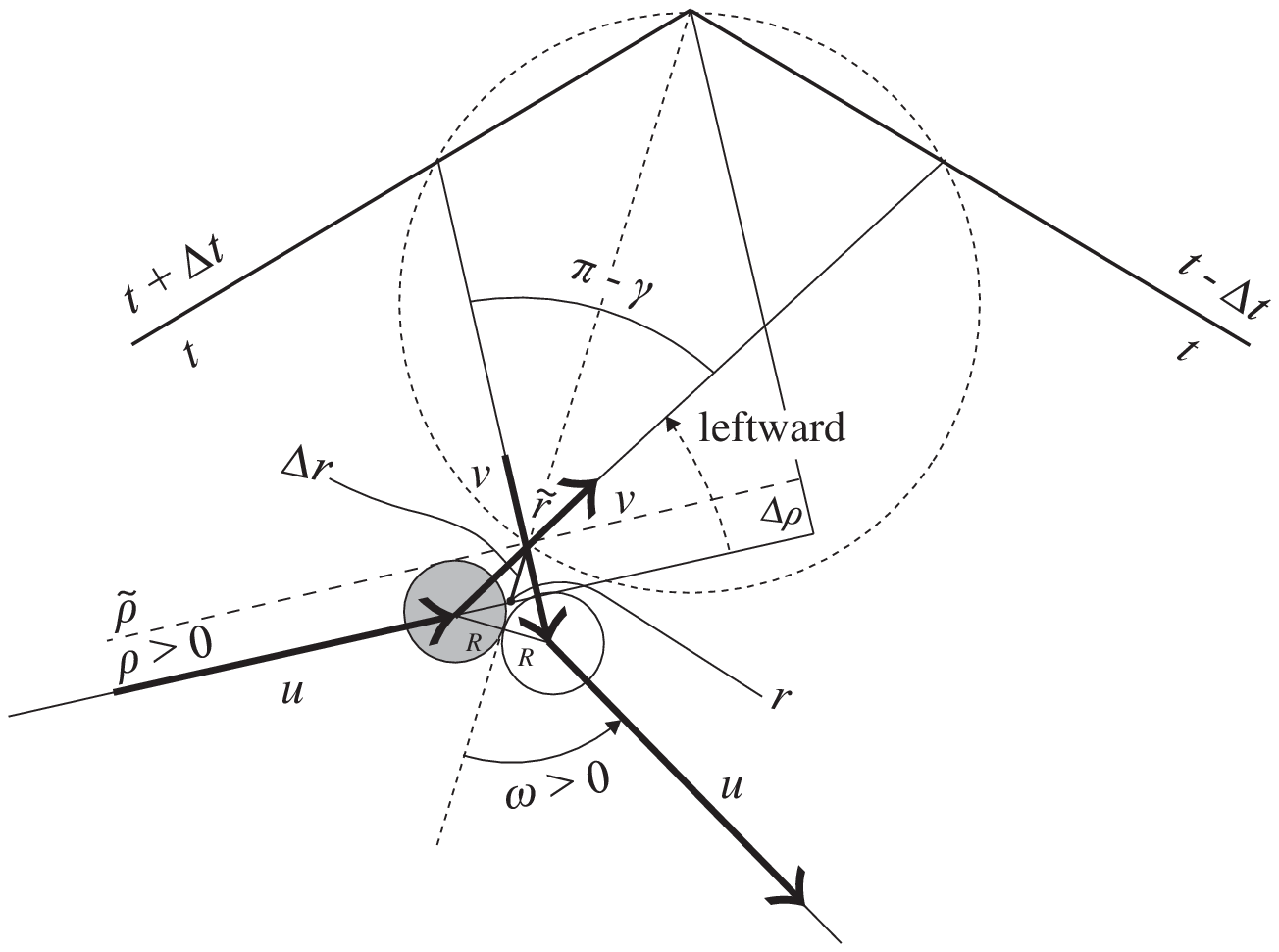}\\[-2ex]
{\small (a)}\\[3ex]
\includegraphics[width=3.3in]{\figdir 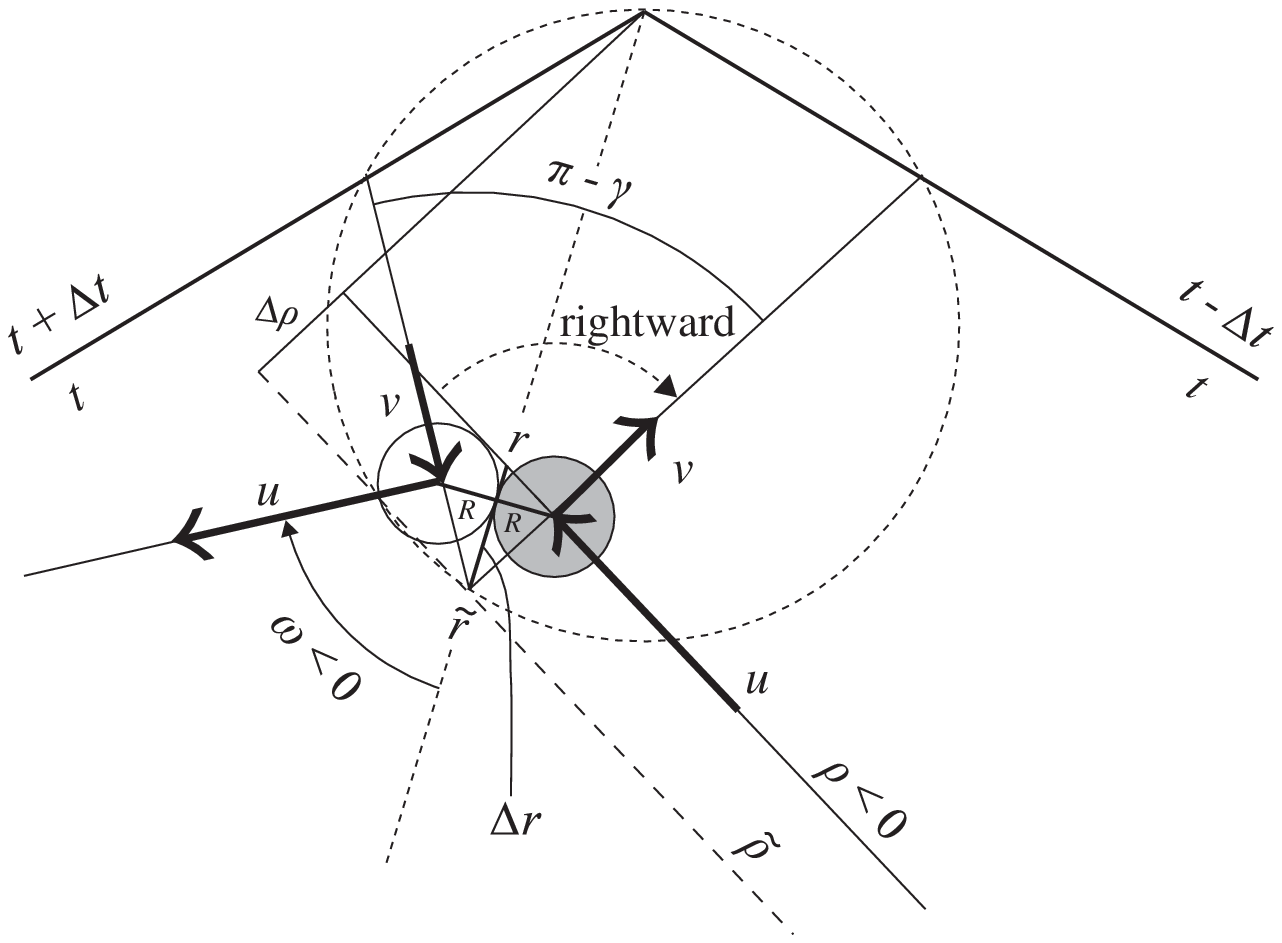}\\[-2ex]
{\small (b)}\\[1ex]
\caption{\label{fig:bCollisions}\small%
\textbf{Physical self-collisions of types I and II.}
(a) The configuration of type I represents situation when the younger ball (in grey) is on the left side while the older ball (in white) is on the right side with respect to the radial line. The self-collision is physical when the momentum transfer is positive and the trajectory of the particle is deflected leftward. (b) The configuration of type II represents situation when the older ball is on the left side while the younger ball is on the right side with respect to the radial line. The self-collision is physical if the trajectory of the particle is deflected rightward.}%
\end{center}\end{figure}
We assume that an elastic collision occurs when the ball collides with itself. The classical elastic impact of two balls is determined by the momentum and energy conservation and the assumption that it occurs in one plane. In our case two versions of the same ball collide: the \emph{younger} version of the ball coming from infinity hits the \emph{older} version coming from the wormhole; after the collision the younger ball flies to the wormhole and the older one away to infinity. Since the traverse of the wormhole does not change the velocity of the ball, magnitude ${v}$ of the outgoing velocity ${\vec{v}}$ of the younger ball is the same as the magnitude of the incoming velocity ${\vec{v}'}$ of the older ball. All these conditions determine \cite{Dolansky:thesis} that the impact of the ball with itself must have the form depicted in Fig.~\ref{fig:impact}. Namely, in the rest frame of the wormhole the incoming velocity ${\vec{u}}$ of the younger ball is complementary to the mirror reflection of the outgoing velocity ${\vec{u}'}$ of the older ball with respect to the plane tangent to balls at the point of impact. We call this plane the \emph{impact plane}. Similar property holds for velocities ${\vec{v}}$ and ${\vec{v}'}$. Moreover, projections of the incoming and outgoing velocities ${\vec{u}}$ and ${\vec{v}}$ (respectively ${\vec{v}'}$ and ${\vec{u}'}$) to the impact plane are the same.

\begin{figure}\begin{center}
\includegraphics[width=3.3in]{\figdir 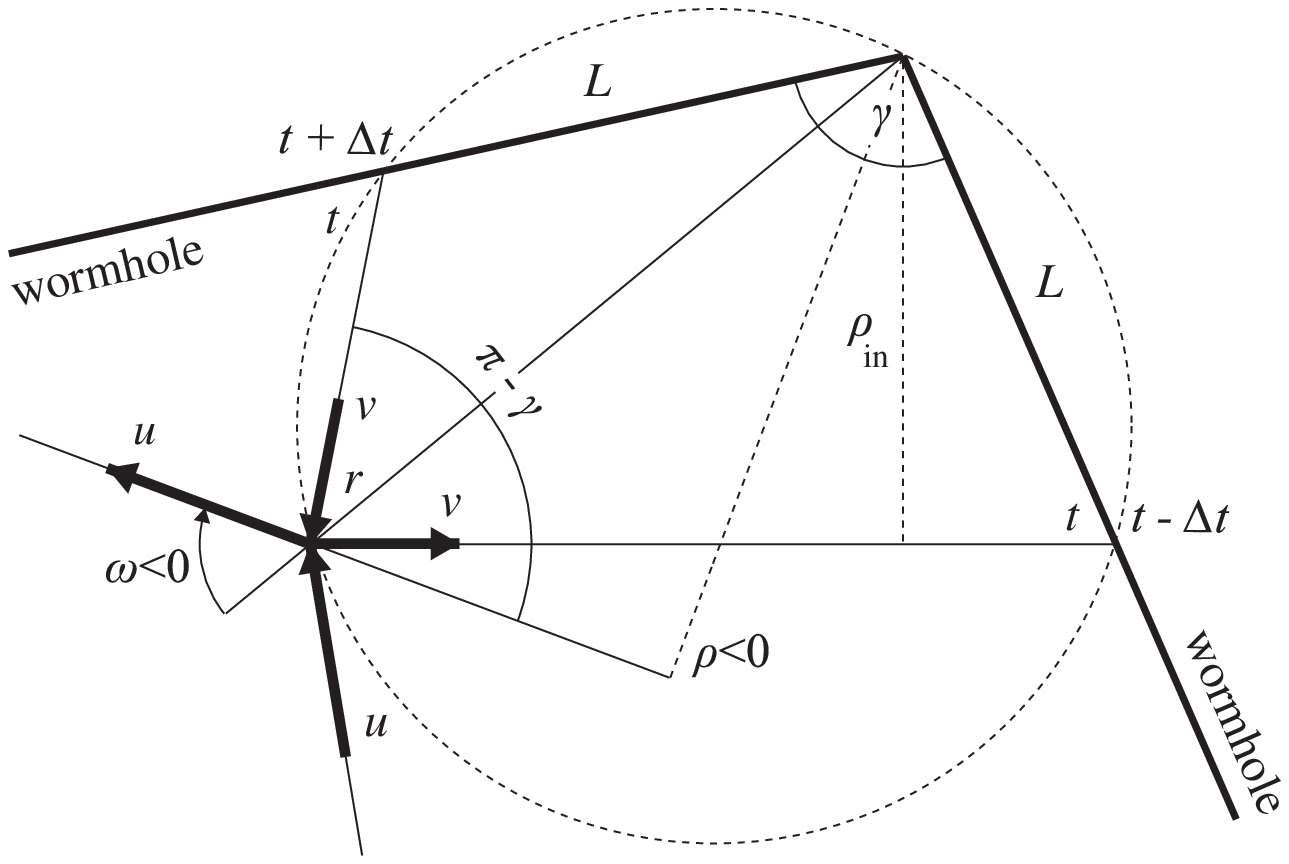}\\[0ex]{\small(a)}\\[3ex]
\includegraphics[width=3.3in]{\figdir 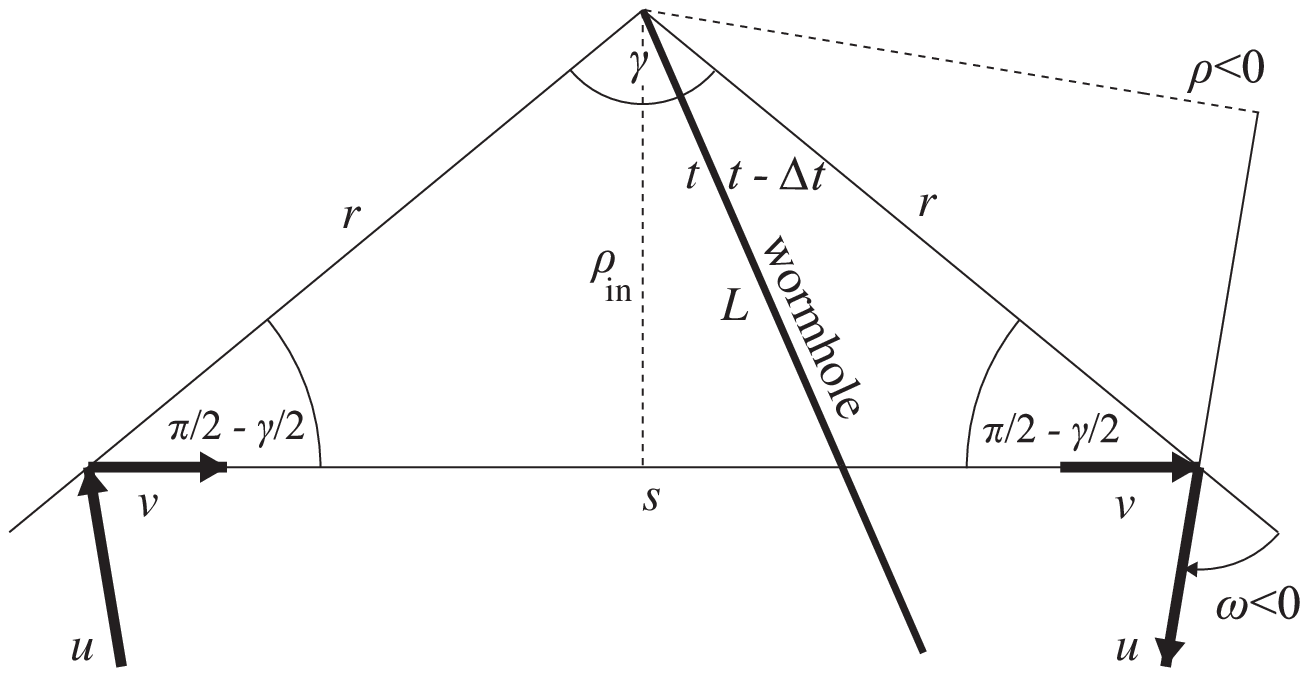}\\[0ex]{\small(b)}\\
\caption{\label{fig:model-parameters}\small%
\textbf{Geometry of self-colliding point-particle trajectory.}
(a) A point particle is approaching the wormhole from infinity with a velocity~${u}$ and impact parameter~${\rho}$, it collides with the version of itself which already passed through the time machine, and with a velocity $v$ moves toward the wormhole. After passing it and self-colliding with itself, it moves with the velocity $u$ back to infinity. The self-collision occurs at the distance~${r}$ from the axis and the outgoing trajectory forms with the radial direction the orientated angle~${\omega}$. Thanks to symmetry of the self-collision with respect to radial direction, the angle~${\omega}$ is a half of the angle between the incoming and outgoing trajectory. (b) The same situation depicted in a map which cuts the conical space not along the wormhole but along the radial direction through the self-collision. The length ${s}$ of the straight trajectory between its self-intersection is given by ${s=2\rho\tan\frac{\gamma}{2}=2r\sin\frac\gamma2}$.}%
\end{center}\end{figure}

Now we want to find a location of the self-collision in such a way that the outgoing younger ball consistently passes to the incoming older ball. It can be done \cite{Dolansky:thesis} in two ways depicted in Fig.~\ref{fig:bCollisions}. The key property is that the impact plane must be radial, i.e., it contains the axis (dashed line in the figure). Type I represents the situation when the older ball touches the younger one by its rear part, i.e., the younger version must collide from the left side.\footnote{%
Left and right side is selected by our choice of time-shift of the time machine.} %
For type II the younger ball touches its older version by its front part and is incoming from the right side.

For both types we can distinguish the physical collisions with a positive momentum transfer from the younger ball to the older ball from ``spurious'' collisions which would need a negative momentum transfer. The latter are unphysical not only for the sign of the exchanged momentum but also because the ball would not fit geometrically into space for such collisions---it would have to fly through itself. For the physical self-collision of type I the younger ball must deflect leftward, i.e., closer to the vertex than if it followed the collision-free trajectory. For the physical self-collisions of type II the trajectory of the younger ball is deflected rightward, cf.\ Fig.~\ref{fig:bCollisions}.

\subsection*{Point-like particle}

To determine parameters of the self-collision, we start with a simpler case of a point particle, i.e., taking radius of the ball ${R\to 0}$. In this case the distinction between the two types described above disappears\footnote{%
More precisely, the physical solutions of type I are geometrically identical to spurious solutions of type II and vice versa. Of course, for a point particle we cannot distinct from which side the particle hits itself.} %
and the geometry of the self-collision is depicted in Fig.~\ref{fig:model-parameters}. The trajectory of the incoming particle can be determined by two initial parameters: the \emph{impact parameter} $\rho$ which gives the distance of the incoming trajectory from the vertex of the cone, and the \emph{magnitude of incoming velocity} ${u>0}$. We adopt the convention that the impact parameter $\rho$ is positive if the particle circles the cone in the counterclockwise direction and it is negative if it circles the cone in the clockwise direction.

Of course, the parameters $u$ and $\rho$ do not determine the incoming trajectory uniquely since they do not specify its angular location around the axis and its temporal location---we would need additional two parameters for that.\footnote{%
Here we completely ignore motion along the axis of the wormhole.} %
However, the conical space has the rotational symmetry and it is static, so the exact angular and temporal location is irrelevant for the character of solutions. Therefore we can ignore the additional initial parameters when we investigate the geometry of the self-collision.

The self-collision can be parametrized by its \emph{radial distance}~${r>0}$ from the axis and by the oriented \emph{scattering angle} ${\omega\in(-\frac\pi2,\frac\pi2)}$ between the outgoing trajectory and the radial direction, see Fig.~\ref{fig:model-parameters}a.

Thanks to the symmetry of the self-collision with respect to the radial direction, the angle ${\omega}$ also determines the direction of incoming trajectory. For the same reason, the outgoing particle also has velocity ${u}$ and impact parameter ${\rho}$. The angle between the inner trajectory and the radial direction is determined by the conical geometry and it is equal to ${\frac\pi2-\frac\gamma2}$, cf.\ Fig.~\ref{fig:model-parameters}b. The length ${s}$ of the inner trajectory is then ${s=2r\sin\gamma/2}$.

The collision parameters ${r}$ and ${\omega}$ encode the same information as the initial parameters ${u}$ and ${\rho}$. Indeed, ${\rho}$ in terms of ${r}$ and ${\omega}$ is given by the simple geometry
\begin{equation}\label{eq:p-rrhorel}
  \rho = r\,\sin\omega\;.
\end{equation}
Since the radial projection of the particle velocity before and after the collision is the same (as a consequence of the laws of the elastic impact as we discussed for finite balls), the incoming velocity ${u}$ is related to the inner velocity ${v}$ along the trajectory between the self-collision as
\begin{equation}\label{eq:p-eq-mot1}
u\cos\omega=v\sin\frac{\gamma}{2}\;.
\end{equation}
However, the inner velocity must be such that the particle passes the inner trajectory exactly in the time ${\Delta t}$ gained in the wormhole,
\begin{equation}\label{eq:p-innervel}
  v = \frac{s}{\Delta t} = \frac{2r\sin\frac\gamma2}{\Delta t}
\end{equation}
We thus obtain relation for ${u}$:
\begin{equation}\label{eq:p-urel}
  u = \frac{2 r \sin^{2}\frac{\gamma}{2}}{\Delta t\cos\omega}\;.
\end{equation}
It will be also useful to write down the relation for ${\rho}$ as function of ${\omega}$ and ${u}$
\begin{equation}\label{eq:rho-parametric}
  \rho=\frac{u\Delta t}{4\sin^{2}\frac{\gamma}{2}} \sin(2\omega)\;.
\end{equation}

Equations \eqref{eq:p-rrhorel} and \eqref{eq:p-urel} thus relate the parameters of the incoming trajectory ${u}$, ${\rho}$ and the parameters of the self-collision ${r}$, ${\omega}$. Their consequences will be discussed in more detail in Sec.~\ref{sc:ppart}.

Before we return to the case of finite balls, let us note that we can introduce similar parameters also for the free trajectory. The only difference is that ${r}$ and ${\omega}$ refer, in general, to self-intersection (i.e., not necessarily to self-collision) of the free trajectory. Clearly, the angle ${\omega}$ is now given just by the conical geometry
\begin{equation}\label{eq:p-omegafree}
  \omega=\omega_{\fr}\equiv\frac{\pi-\gamma}{2}\;.
\end{equation}
The solutions, for which the younger version of the particle passes through the point of self-intersection later than the older version (coming from the time machine), are the solutions of type~I. When the younger particle passes the point of self-intersection earlier, we speak about the solutions of type~II.

The self-intersection threatens to become a self-collision if time spent on the inner part of the free trajectory is equal to the time-shift of the time machine. Since for a free trajectory ${u=v}$, conditions \eqref{eq:p-innervel} and \eqref{eq:p-rrhorel} give
\begin{equation}\label{eq:p-rhopx}
u_{\px}\,\Delta t =  2\rho_{\px}\tan\frac\gamma2\;.
\end{equation}
Such values ${u_{\px}}$, ${\rho_{\px}}$ correspond to the ``paradoxical'' free trajectory for which the particle occurs at the point of self-intersection twice at the same time. For a point particle, it is not possible to decide plausibly what happens in such a situation. We can only observe, that the same parameters ${u_{\px}}$, ${\rho_{\px}}$ also describe the trajectory with one self-collision, which is given by the same parameters as ``paradoxical'' self-intersection of the free trajectory, namely ${r=\rho_{\px}/\cos\frac\gamma2}$ and ${\omega=\omega_{\fr}}$. A detailed analysis of this situation will be done for finite balls in Sec.~\ref{sc:numbsol}.

The direction ${\omega_{\fr}}$ also plays a role of the limiting value between physical collisions of type I and II. Physical self-collisions of type I, for which the trajectory deflects leftward from the free trajectory, are bounded within the interval ${\omega\in(\omega_{\fr},\frac\pi2)}$; while physical self-collisions of type II (with rightward deflection) sweep out the interval ${\omega\in(-\frac\pi2,\omega_{\fr})}$.

\subsection*{Finite ball}

We have to be more precise to define the parameters of the self-collision for a finite ball. Let ${r}$ be the radial distance from the axis of the intersection of the incoming and outgoing trajectories (extended beyond the actual self-collision), and ${\tilde{r}}$ be the radial distance from the axis of the self-intersection of the inner trajectory (for type II extended beyond the self-intersection), see Fig.~\ref{fig:bCollisions}. Clearly, Eq.~\eqref{eq:p-rrhorel} still holds, and
${r=\tilde{r} \pm R(\tan\gamma/2-\cot\omega)}$,
with upper sign for type I and lower sign for type II.
The geometry of the inner trajectory of the finite balls is identical with that of a point particle with modified impact parameter ${\tilde\rho=\tilde r \sin\omega}$. The path ${s}$ traveled during time ${\Delta t}$ gained in the time machine must be corrected due to finite radius of balls,
${s=\tilde s \pm \frac{2R}{\cos\gamma/2}}$,
where, analogously to the point-particle case, ${\tilde s=2\tilde r\sin\gamma/2}$ is the length of the inner trajectory between its self-intersection. The corrected relation \eqref{eq:p-innervel} together with \eqref{eq:p-eq-mot1} finally leads to equation between ${u}$, ${\rho}$, and ${\omega}$:
\begin{equation}\label{eq:b.omegaurho}
  \rho=\frac{1}{4\sin^{2}\frac{\gamma}{2}}
    \Bigl(u\triangle t \sin2\omega\mp 4R\sin\frac{\gamma}{2}\sin\bigl(\frac{\gamma}{2}+\omega\bigr)\Bigr)\;,
\end{equation}
again, with upper/lower sign for type~I/II, respectively.

Physical self-collisions of type I or II are restricted by the conditions
\begin{equation}\label{eq:b.type}
\begin{aligned}
    \omega&>\omega_{\fr}\qquad &&\text{for physical solutions of type I}\;,\\
    \omega&<\omega_{\fr}\qquad &&\text{for physical solutions of type II}\;,\\
\end{aligned}
\end{equation}
otherwise the ball would deflect to a wrong side of the free trajectory and the momentum transfer from the younger to older version of the ball would be negative.

Eq.~\eqref{eq:b.omegaurho} supersedes relation~\eqref{eq:rho-parametric} for the point particle, as can be also seen by taking limit ${R\to0}$. This can be regarded as the key equation of motion which determines the scattering angle ${\omega}$ in terms of the initial conditions ${u}$ and ${\rho}$. The position of the self-collision is then determined by the parameter ${r}$ which is given by Eq.~\eqref{eq:p-rrhorel}.

\section{Number and character of solutions}
\label{sc:numbsol}

\subsection*{Point-like particle}

\begin{figure}\begin{center}
\includegraphics[width=3.3in]{\figdir 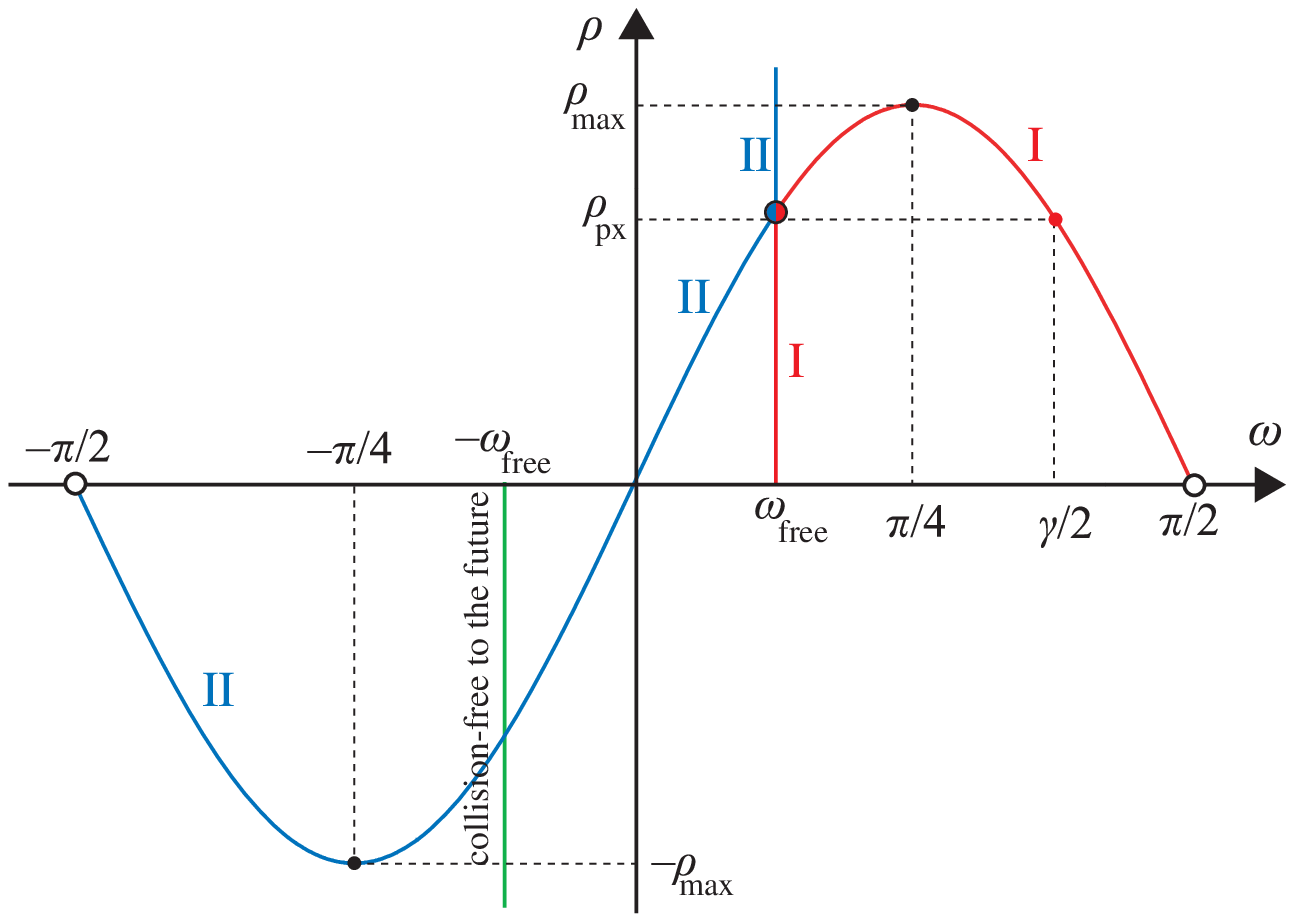}\\[3ex]
\caption{\label{fig:p-rho(omega)}\small%
\textbf{The $\boldsymbol{\rho$-$\omega}$ diagram of consistent solutions for a point particle.}
Each point of the graph in the ${\rho}$-${\omega}$ plane represents a consistent solution for a point particle with at most one self-collision. The velocity ${u}$ is fixed (it changes only a scale of the solutions) and the parameter ${r}$ is given by Eq.~\eqref{eq:p-rrhorel}.
Collision-free solutions are represented by the straight vertical lines. The line ${\omega=-\omega_{\fr}}$ (in green) represents collision-free trajectories to the future, the line ${\omega=\omega_{\fr}}$ represents free trajectories to the past. The points below the value ${\rho_\px}$ (in red) corresponds to the solutions of type I, for which the younger particle passes the point of self-intersection later than that coming from the wormhole. The solutions with ${\rho>\rho_\px}$ (in blue) are of type II.
The sinusoidal part of the graph represents solutions with one self-collision. The part with ${\omega>\omega_{\fr}}$ (in red) represents self-collisions of the type~I, the part with ${\omega<\omega_{\fr}}$ (in blue) represent self-collisions of type II.
Self-colliding solutions are thus possible only for ${|\rho|<\rho_{\mx}}$.
The point ${(\rho_\px,\,\omega_{\fr})}$ represents both the paradoxical collision-free solutions for which younger and older versions of the particle meet at the point of intersection at the same moment, and the consistent self-colliding solution (which is geometrically identical to the paradoxical solutions). Exact relation of these solution is clarified in the case of balls with a finite radius.}\vspace*{-3ex}
\end{center}\end{figure}

In the case of the point particle there is only one length scale ${u \Delta t}$ given by the initial velocity~${u}$ and the time-shift ${\Delta t}$. The velocity ${u}$ thus changes only the scale of the whole experiment. Therefore, we can fix ${u}$ and study only the relations among ${\rho}$, ${r}$, and ${\omega}$.

The relation between ${\omega}$ and ${\rho}$ is given by Eq.~\eqref{eq:rho-parametric} and depicted in Fig.~\ref{fig:p-rho(omega)}, the relation to the parameter ${r}$ is given by \eqref{eq:p-rrhorel} (or, alternatively, implicitly by Eq.~\eqref{eq:p-urel}).

We immediately see that the self-collision can happen only for \begin{equation}\label{eq:rhomax}
  |\rho|<\rho_{\mx}\equiv\frac{u\Delta t}{4\sin^2\frac\gamma2}\;.
\end{equation}
For larger ${|\rho|}$ the particle would not be fast enough to travel through the wormhole and hit itself consistently---it would be too far from the axis of the wormhole and the path through the wormhole would be too long.

\begin{figure}\begin{center}
\includegraphics[width=3.3in]{\figdir 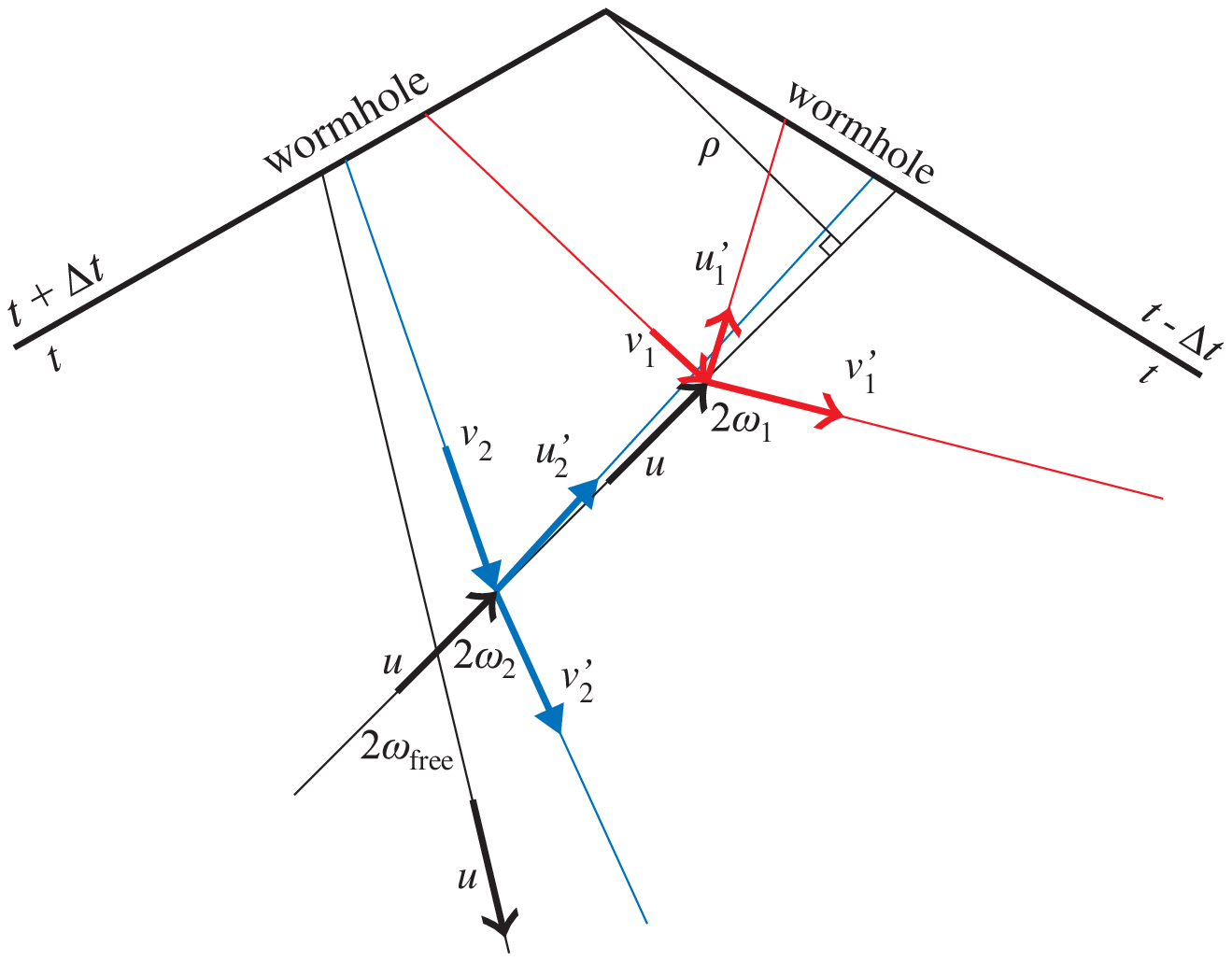}
\caption{\label{fig:p.two.solutions}\small%
\textbf{Three possible solutions for given initial conditions of a point-like particle.}
For ${|\rho|<\rho_{\mx}}$, ${\rho\neq\rho_\px}$ there exist three possible solutions: the collision-free trajectory (black) and two self-colliding trajectories (red/gray and blue/light-gray). One of the self-colliding trajectories is close to the free trajectory (it coincides in the limit ${\rho\to\rho_{\px}}$), another is rather different.}\vspace*{-3ex}
\end{center}\end{figure}

For ${|\rho|<\rho_{\mx}}$, the map ${\rho\to\omega}$ is not a unique relation: for a given ${\rho}$ we have, in general, two ${\omega}$ satisfying \eqref{eq:rho-parametric}. For the given initial conditions, we thus obtain, in addition to the collision-free trajectory, two self-colliding solutions. These two solutions have the scattering angles complementary to ${\pm\frac\pi2}$ and represent completely different evolutions of the system. Note, that the self-colliding solutions exist even if ${\rho<0}$, i.e., if the free trajectory passes the wormhole to the future.

All three possible solutions for the given initial conditions (with ${\rho<\rho_{\mx}}$) are shown in Fig.~\ref{fig:p.two.solutions}.

For the paradoxical value ${\rho=\rho_{\px}}$, Eq.~\eqref{eq:p-rhopx}, the ``pardoxical'' collision-free solution geometrically coincides with one of the self-colliding solutions. It is called ``paradoxical'', since for this initial values the free moving particle meets itself at the point of self-intersection exactly at the same moment, so it cannot be the collision-free solution. However, since for a point-like particle we cannot distinguish the non-colliding solution from the colliding one, it is dubious to study the nature of this ``paradoxical'' situation. Therefore, we will discuss the details of the ``paradoxical'' initial conditions in the case of finite balls. After that we will return to the simpler case of a point-like particle to study the geometry of the trajectories in more detail.

\subsection*{``Paradoxical'' situations for a finite ball}

In the case of finite balls the system has an additional length scale given by the radius ${R}$ of the ball. Therefore, the dependence on the velocity ${u}$ is not trivial any more. Nevertheless, we still fix the value of the initial velocity and discuss the structure of the corresponding solutions. The values of various quantities can, however, depend on ${u}$ in a more complicated way than just a rescaling.

\begin{figure}\begin{center}
\includegraphics[width=3.3in]{\figdir 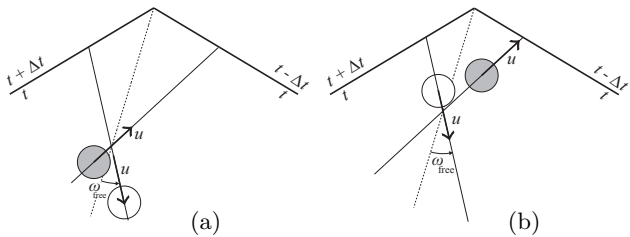}
\\[-2ex]
{\small \hspace*{0.5in}(a)\hspace{1.5in}(b)}
\caption{\label{fig:collision-free}\small%
\textbf{Two types of collision-free trajectories passing the time machine to the past.}
(a) Trajectory of type~I. For ${\rho<\rho_\pxI}$, time ${s/v}$ spent by the ball between the self-collision is smaller than the time-shift ${\Delta t}$, i.e., the younger version of the ball goes through the point of self-intersection later than the older version. (b)~Trajectory of type~II. For ${\rho>\rho_\pxII}$ the younger version of the ball goes through the point of intersection earlier than the older version.}\vspace*{-3ex}
\end{center}\end{figure}

Collision-free trajectories of a finite ball are again characterized by the angle ${\omega_\fr}$ and they can be also divided into types I and II, see Fig.~\ref{fig:collision-free}. However, due to the finite radius of the balls, there is a whole interval of the impact parameter ${\rho}$, for which the collision-free trajectory becomes ``paradoxical''---if the ball hits itself inconsistently. Checking the geometry of the free trajectory of a ball, it turns out \cite{Dolansky:thesis}, that the boundary of this interval is given by values of ${\rho}$ determined by Eqs.~\eqref{eq:b.omegaurho} for types I and II with ${\omega=\omega_{\fr}}$,
\begin{equation}\label{eq:b.pxrho}
    \text{${\rho}$ paradoxical}\quad\Leftrightarrow\quad
    \rho\in(\rho_\pxI,\rho_\pxII)\;.
\end{equation}
Explicitly, the ``paradoxical'' values of the impact parameter are given by
\begin{equation}
\rho_{\mathrm{pxI,II}}=\frac{u\triangle t}{2\tan\frac{\gamma}{2}}\mp\frac{R}{\sin\frac{\gamma}{2}}\;.
\end{equation}
For ${\rho<\rho_\pxI}$ the collision-free trajectory is of type I, i.e., the older version of the ball overtakes the younger one at the point of intersection. If ${\rho>\rho_\pxII}$, the collision-free trajectory is of type II, cf.\ Fig.~\ref{fig:collision-free}.

\begin{figure}\begin{center}
\includegraphics[width=3.3in]{\figdir 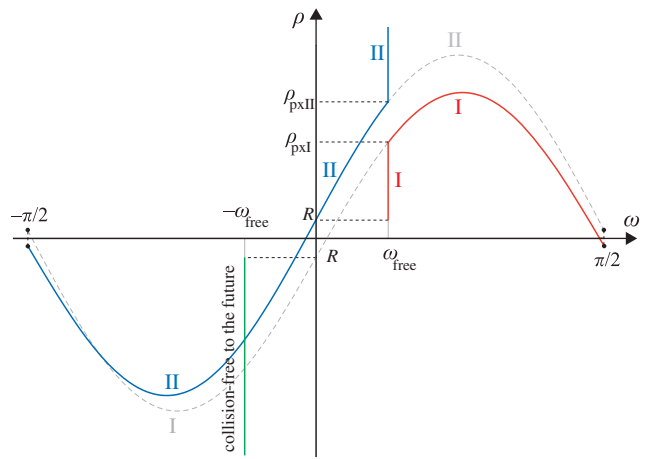}
\caption{\label{fig:rho&r(omega)-a}\small%
\textbf{Physical solutions in the $\boldsymbol{\rho$-$\omega}$ plane.}
The points on the curves represent solutions characterized by the parameters ${\rho}$ and ${\omega}$. The initial velocity ${u}$ is fixed. The physical solutions correspond to the solid curves, the spurious one to the dashed curves. Vertical lines are collision-free solutions. Sinusoidal curves, determined by Eqs.~\eqref{eq:b.omegaurho}, represent the solution with one self-collision.
The type of the solution is indicated in a similar way as in Fig.~\ref{fig:p-rho(omega)}.}
\end{center}\end{figure}

The parameters of the self-colliding trajectories are characterized by relations \eqref{eq:b.omegaurho} and \eqref{eq:p-rrhorel}. We can depict the physical solutions (cf.\ conditions~\eqref{eq:b.type}) with at most one self-collision in a diagram analogous to the one we used in the point-particle case, see Fig.~\ref{fig:rho&r(omega)-a}.

We see here that the solutions split into two branches: the physical self-colliding solutions of type~I together with collision-free solutions of type~I, and self-colliding and collision-free solutions of type~II. The collision-free solution of type~I becomes self-colliding solution of type~I for ${\omega=\omega_\fr}$, ${\rho=\rho_\pxI}$. It is limiting case when the ball on the free trajectory just touches itself but it does not exchange any momentum. It is thus identical to the limiting case of physical self-colliding solutions of type~I. Similarly, for ${\omega=\omega_\fr}$, ${\rho=\rho_\pxII}$ the ball on the free trajectory just touches itself, now from the other side, and it becomes the self-colliding solution of type II.

\begin{figure}\begin{center}
\includegraphics[width=3.3in]{\figdir 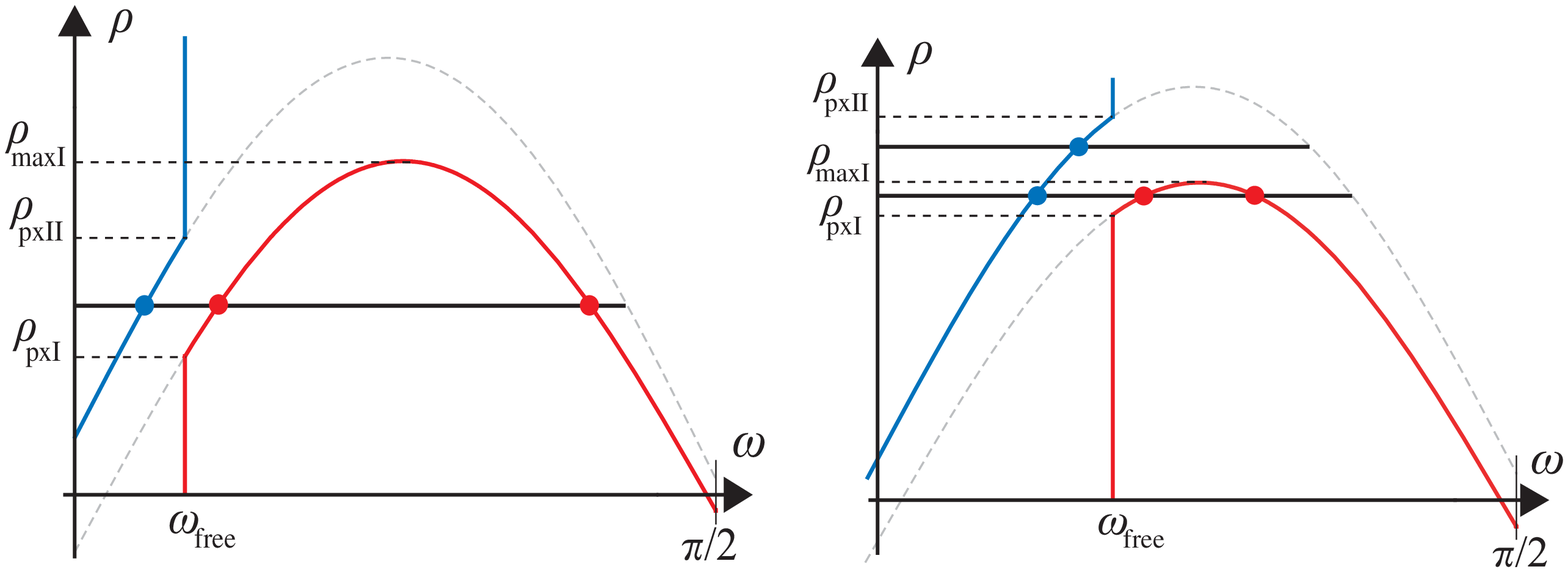}
{\small (a)\hspace{1.65in}(b)}
\caption{\label{fig:ball-structure-dangerous}\small%
\textbf{Number of solutions for given initial conditions.}
The number of solutions for chosen initial parameter ${\rho}$ can be determined from Fig.~\ref{fig:rho&r(omega)-a} by intersecting the graph with the horizontal line corresponding to ${\rho}$. The choice of the initial velocity ${u}$ can slightly modify a shape of the diagrams. Here, two representative cases are shown. Only the part with ${\omega>0}$ is depicted---it approximately corresponds to the initial conditions ${\rho>0}$ with the ball directed to the time machine toward the past.
Clearly, the number of solutions steps from one to three with ${\rho}$ becoming smaller than ${\rho_\mxI}$. In the ``paradoxical'' interval ${(\rho_\pxI,\rho_\pxII)}$ the number of solutions remains the same. Only the character of the solutions is different: the collision-free solution is superseded by the self-colliding solution. In the case (b) the limiting value ${\rho_\mxI}$ belongs to the ``paradoxical'' interval. For ${\rho\in(\rho_\mxI,\rho_\pxII)}$ there exists just one self-colliding solution.}\vspace*{-3ex}
\end{center}\end{figure}

Inspecting Fig.~\ref{fig:rho&r(omega)-a}, we can also conclude, that for the initial conditions with ${\rho\not\in(-\rho_\mxII,\rho_\mxI)}$ (where ${\rho_{\mathrm{maxI,II}}}$ are given by the maximal values of the expression \eqref{eq:b.omegaurho} for type I and II, respectively), there exist just one consistent solution. It is typically a collision-free trajectory, however, for certain values of parameters it can also be a self-colliding solution, cf.\ Fig.~\ref{fig:ball-structure-dangerous}b.

\begin{figure}[b]\begin{center}
\includegraphics[width=3.3in]{\figdir 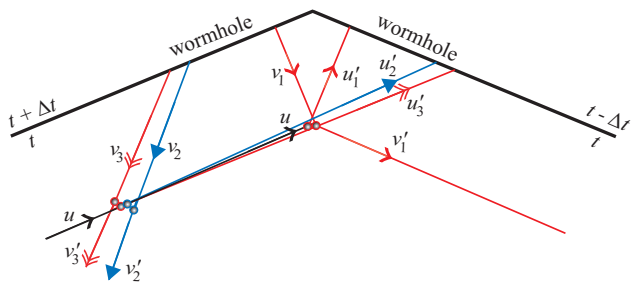}
\caption{\label{fig:bDangerousRegion}\small%
\textbf{Three self-colliding solutions for $\boldsymbol{\rho}$ from the ``paradoxical'' interval.}
Unlike the point-like case, there is a whole interval ${(\rho_\pxI,\rho_\pxII)}$, where collision-free trajectories do not exist. As we can see from Fig.~\ref{fig:ball-structure-dangerous}a, the collision-free trajectory is replaced by a self-colliding one. The diagram depicts all three self-colliding solutions, two of type I (in light and dark red) and one of type II (in blue).}\vspace*{-3ex}
\end{center}\end{figure}

For initial conditions with ${\rho\in(-\rho_\mxII,\rho_\mxI)}$, there are three possible solutions; typically, one collision-free and two self-colliding. However, for the paradoxical values of the impact parameter ${\rho\in(\rho_\pxI,\rho_\pxII)}$, the collision-free solution is superseded by a self-colliding solution, cf.\ Fig.~\ref{fig:ball-structure-dangerous}. In this case, we obtain two self-colliding solutions of type~I and one solution of type~II, see~Fig.~\ref{fig:bDangerousRegion}.

We can conclude that the ``paradoxical'' initial conditions are not paradoxical in any dangerous way. These initial conditions lead to the same number of solutions as other sufficiently close initial conditions. Only, the character of solutions is different: the collision-free solution changes to the self-colliding solution. The number of solutions depends on other characteristics, namely if ${\rho\in(-\rho_\mxII,\rho_\mxI)}$: for a large ${|\rho|}$ only the collision-free solution is admissible.

\begin{figure}\begin{center}
\includegraphics[width=3.3in]{\figdir 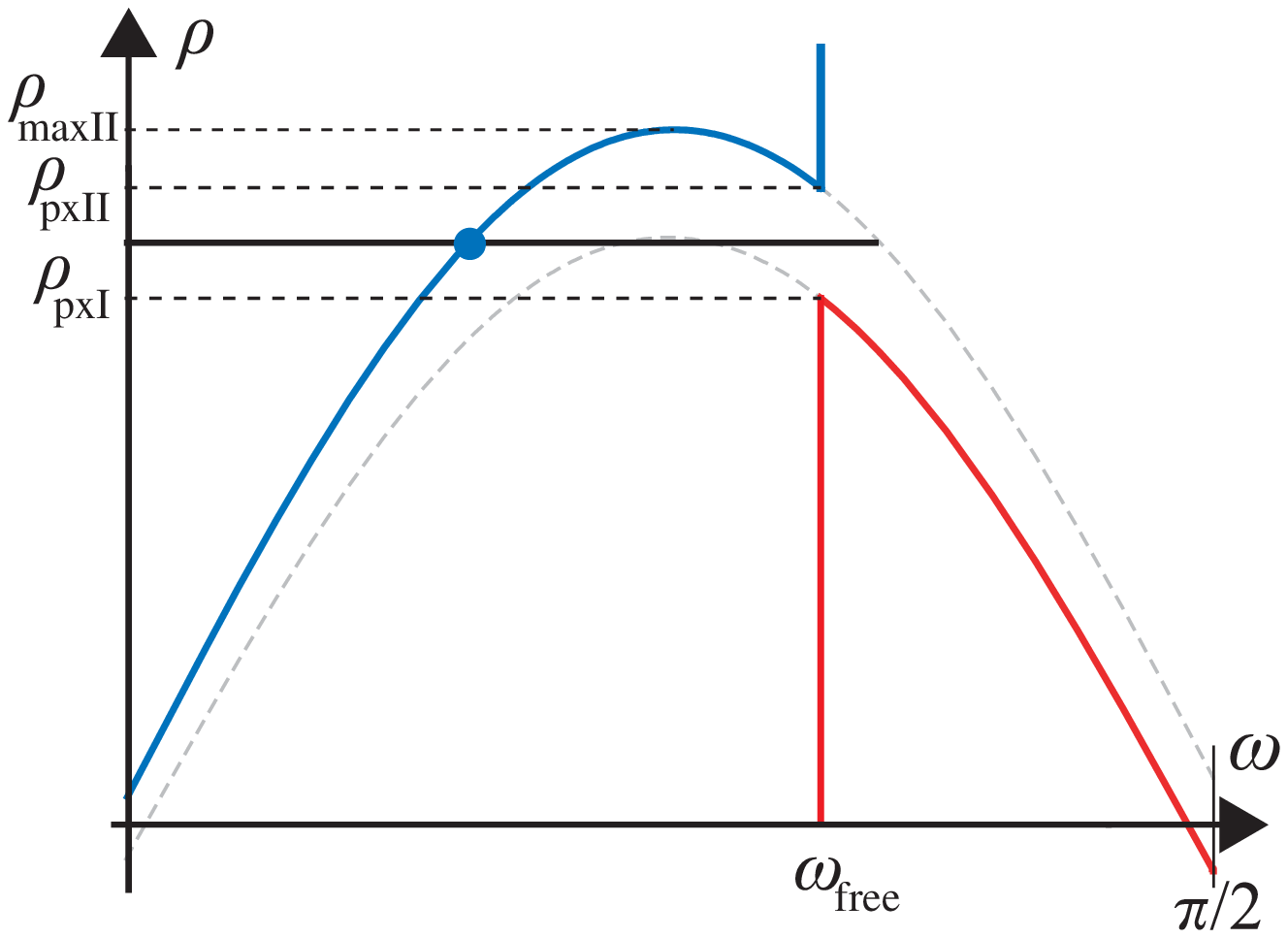}\\[-1ex]
{\small \mbox{}}
\caption{\label{fig:paradox}\small%
\textbf{Solutions from ``paradoxical'' interval for ${\boldsymbol{\gamma<\frac\pi2}}$.}
In this case the collision-free angle ${\omega_\fr>\frac\pi4}$. For the impact parameter ${\rho}$ from the corresponding paradoxical interval ${\rho\in(\rho_\pxI,\rho_\pxII)}$ we find a gap in the solution curves. For a paradoxical ${\rho}$ we obtain only one self-colliding solution. Moreover, this solution corresponds to a value of ${\omega}$ which is substantially different from ${\omega_\fr}$.}\vspace*{-3ex}
\end{center}\end{figure}

However, it should be said that the situation changes if the angle between the wormhole mouths is smaller. For ${\gamma<\frac\pi2}$, the equations for the trajectory remain the same but the value of the angle ${\omega_\fr}$ is larger than ${\frac\pi4}$. The structure of the solutions thus changes as depicted in Fig.~\ref{fig:paradox}. Clearly, for ${\rho\in(\rho_\pxI,\rho_\pxII)}$ the collision-free trajectory is not possible and it is not superseded by another solution. Moreover, for such ${\rho}$, there exists only one self-colliding solution. In this case we can indeed speak about paradoxical initial conditions since the solutions evolved from these conditions are really restricted. However, in the case ${\gamma<\frac\pi2}$ the whole discussion is more complicated since the particle can self-interact in a more complicated way (there is possibility of self-intersection between several self-collisions, etc.). We do not have any indications that these more complicated processes could improve the discussed behavior, however we have not excluded it completely \cite{Dolansky:thesis}. In the following we restrict again to the larger angles, ${\gamma>\frac\pi2}$.



\section{Motion of a point-like particle}
\label{sc:ppart}

Clarifying the character of solutions with ``paradoxical'' initial conditions, in this section we describe the motion of a point-like particle in more detail.

The geometry of the inner trajectory is given only by the parameter ${r}$ and by the angular position of the self-collision with respect to the wormhole---i.e., it is independent of the scattering angle ${\omega}$. The geometry is such that the point of self-collision, the point of the entry to wormhole, the point of the departure from the wormhole, and the axis itself, belong to a common circle, cf.\ Fig.~\ref{fig:model-parameters}a. The inner particle velocity ${v}$ is given by Eq.~\eqref{eq:p-innervel}. For the fixed point of self-collision one can then choose any ${\omega\in(-\frac\pi2,\frac\pi2)}$, which determines ${\rho}$ through Eq.~\eqref{eq:p-rrhorel}, and calculate the initial velocity using Eq.~\eqref{eq:p-eq-mot1}.

Taking into account the rotational symmetry of the conical space between the mouths of the wormhole, we can ignore the angular position with respect to the wormhole in the discussion about many features of the motion. But, when discussing the asymptotic behavior of the incoming and outgoing trajectories, the angular information must be taken into account. Namely, one has to track if the outer trajectories pass through the wormhole. Clearly, it must occur for the scattering angle ${\omega>\frac\gamma2}$, but, depending on the angle between the self-collision and wormhole, it can happen also for smaller angles ${\omega}$.

However, we first describe the motion without a reference to the wormhole mouths. It can be done in the simplest way employing the totally covering space for our conical space. Namely, instead of the conical space with angular coordinate ${\ph\in(-\frac\gamma2,\frac\gamma2)}$, we use the space without any restriction on ${\ph}$, i.e., a helical surface winding around the axis infinitely. The original conical space is then obtained by the identification of points which differ in coordinate ${\ph}$ by an integer multiple of ${\gamma}$.

\begin{figure}\begin{center}
\includegraphics[width=3.3in]{\figdir 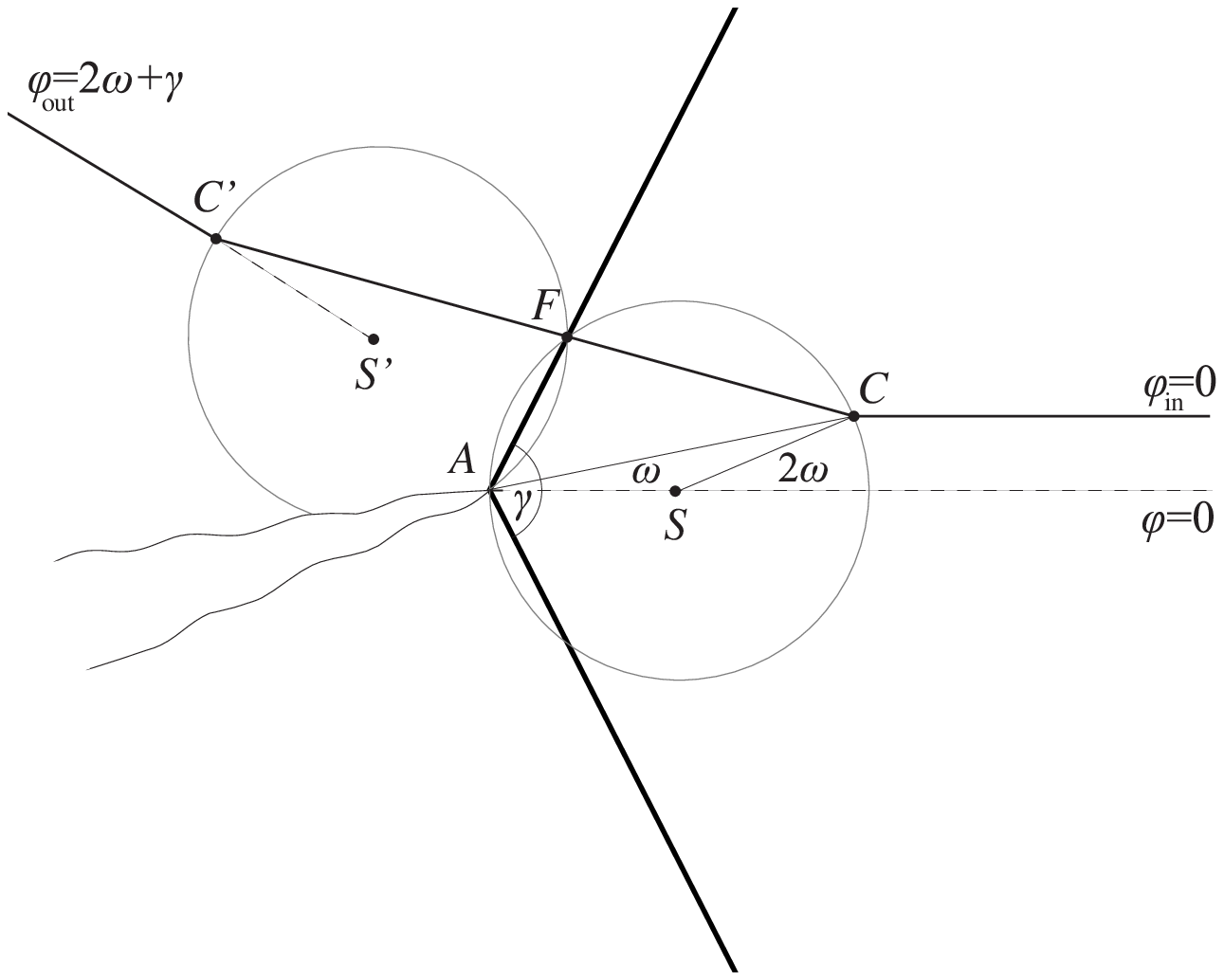}
\caption{\label{fig:trajgeometry}\small%
\textbf{Self-colliding trajectory in the totally covering space.}
The particle incoming along the direction ${\ph_\inc=0}$ with the impact parameter ${\rho}$ is deflected at the point of self-collision ${C}$. The point ${C}$ must lie on the collision circle with the center ${S}$ on the radial line ${\ph=0}$, it passes through the axis ${A}$, and its radius is ${\rho_\mx}$. The trajectory continues toward the point ${C'}$, which is obtained by a contra-clockwise rotation of the point ${C}$ around ${A}$ by angle ${\gamma}$. In the original conical space points ${C}$ and ${C'}$ are identified and correspond to the point of self-collision. From ${C'}$ the trajectory continues in the direction which aims from the center ${S'}$ of the rotated collision circle. The direction of the inner trajectory goes through the focusing point ${F}$. The diagram shows only a part of the totally covering space.}\vspace*{-5ex}
\end{center}\end{figure}

In this covering space the trajectory can be described as follows. Let us assume a particle incoming along the direction ${\ph_\inc=0}$ with the impact parameter ${\rho}$ and the initial velocity ${u}$. The self-collision ${C}$ for such a trajectory always happens on the circle (let us call it the \emph{collision circle}) which has the center ${S}$ on the radial line ${\ph=0}$, it passes through the axis, and its radius is ${\rho_\mx}$, cf.\ Fig.~\ref{fig:trajgeometry}. For ${\rho\in(-\rho_\mx,\rho_\mx)}$, the incoming trajectory intersects this circle twice, which corresponds to two possible self-colliding solutions. The angular coordinate ${\ph}$ of self-collision is given exactly by the scattering angle ${\omega}$. At the point of self-collision ${C}$ the trajectory is deflected, and it continues toward the point ${C'}$, which can be obtained by a contra-clockwise rotation of the point ${C}$ around the axis by angle ${\gamma}$, cf.\ Fig.~\ref{fig:trajgeometry}. In the covering space ${C}$ and ${C'}$ are different points, however, in the original conical space these points are identified as the point of self-collision---of course, the particle must pass the point of self-collision twice. Finally, from ${C'}$ the particle continues through the covering space in the direction which aims from the center ${S'}$ of the rotated collision circle, cf.\ Fig.~\ref{fig:trajgeometry}.

\begin{figure}\begin{center}
\includegraphics[width=3.3in]{\figdir 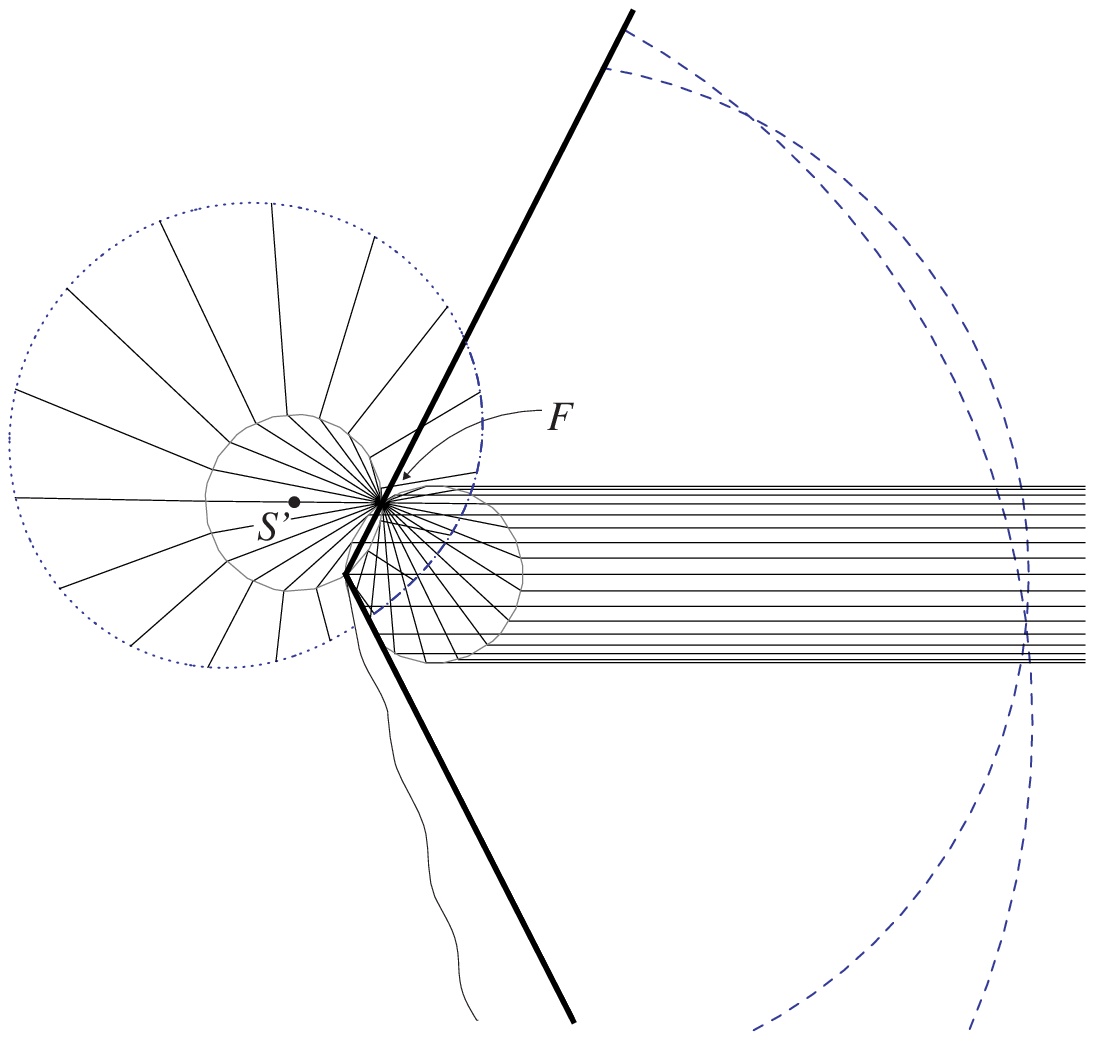}
\caption{\label{fig:congruence}\small%
\textbf{Scattering of the plane-wave congruence of particles in the totally covering space.}
Congruence of particles coming along ${\ph_\inc=0}$ direction in ``plane-wave'' configuration (i.e., aligned at an initial moment on a planar ``wavefront'' perpendicular to the direction of motion) approaches the time machine. The particles scatters on the collision circle and moves toward the corresponding points on the rotated collision circle. Here, they are deflected in the directions coming from the center $S'$ of the rotated collision circle. The dotted curve depicts the wavefront after the scattering in the totally covering space. The particles do not scatter on the collision circle at the same time; however, for large final times the wavefront after scattering approaches the circle. The real wavefront projected back to the conical space is drawn as dashed curves. Segments projected from various sheets of the totally covering space gain additional time-shift thanks to the passage through the time machine. Therefore, these parts of the wavefront are larger since the particle had more time for their motion. Inner trajectories focus at one focusing point, for ${\omega\in(-\frac\gamma2,\frac\gamma2)}$ they pass through this point. The diagram shows only part of the totally covering space, given approximately by ${\ph\in(-\frac\gamma2,2\pi-\frac\gamma2)}$.}\vspace*{-5ex}
\end{center}\end{figure}

\begin{figure*}
\begin{flushleft}
\hspace*{70pt}\includegraphics[scale=0.8]{\figdir 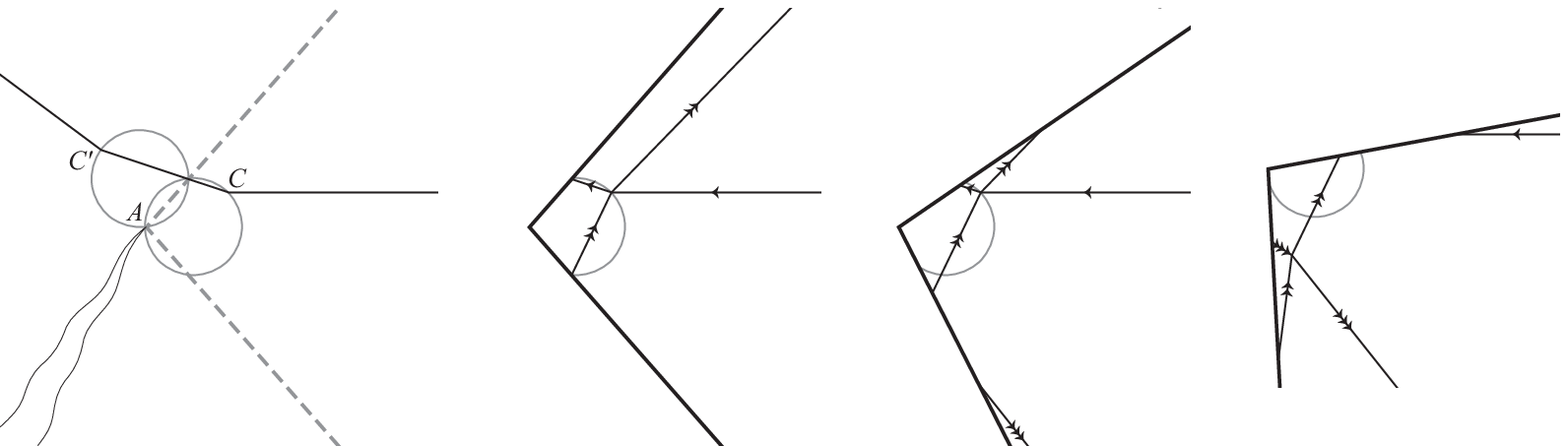}\\[-2ex]
\hspace*{70pt}{\small \qquad\quad(a)}\\[2ex]
\hspace*{70pt}\includegraphics[scale=0.8]{\figdir 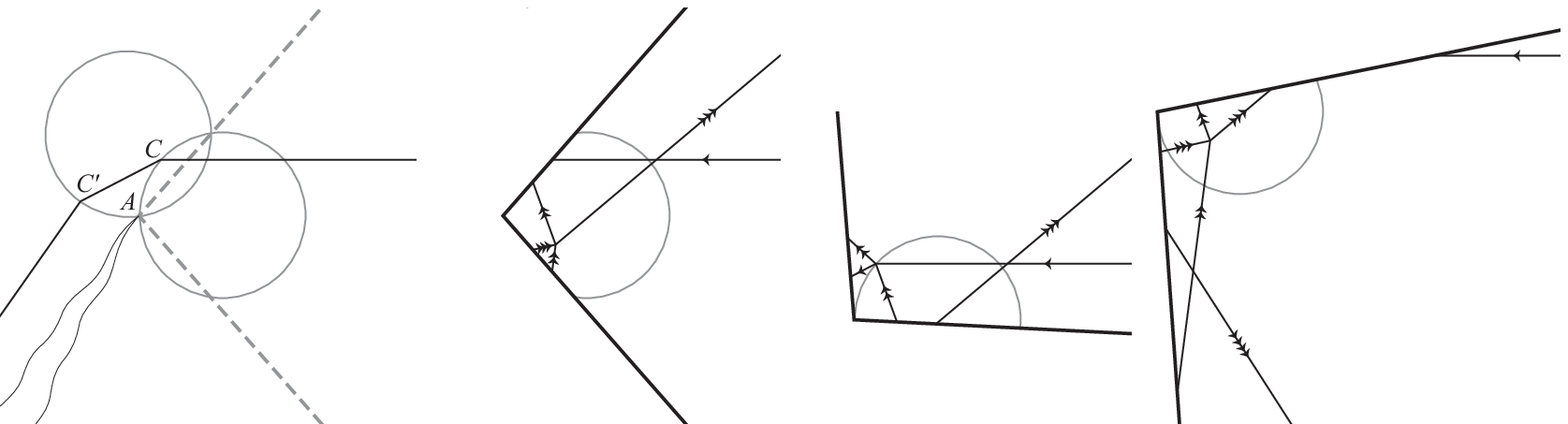}\\[-2ex]
\hspace*{70pt}{\small \qquad\quad(b)}\\[2ex]
\hspace*{70pt}\includegraphics[scale=0.8]{\figdir 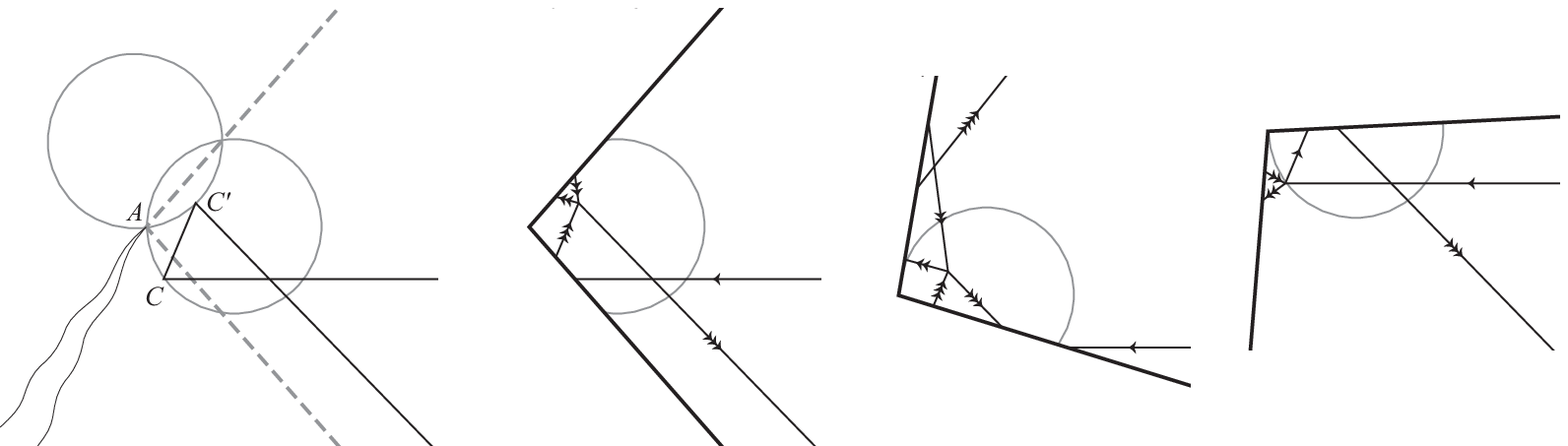}\\[-2ex]
\hspace*{70pt}{\small \qquad\quad(c)}\\[2ex]
\end{flushleft}
\begin{center}
\includegraphics[scale=0.8]{\figdir 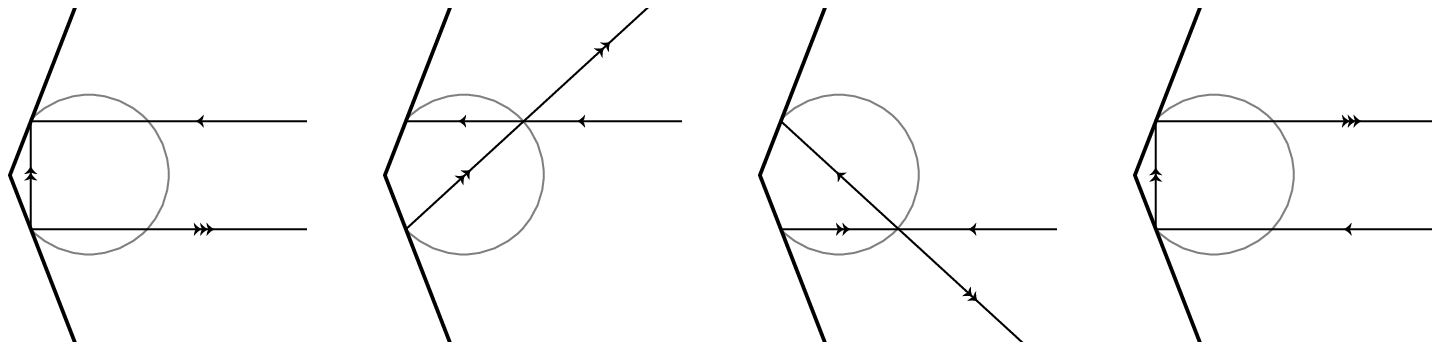}\\[-2ex]
{\small \hspace*{12pt}(d)\hspace{76pt}(e)\hspace{76pt}(f)\hspace{76pt}(g)}\\[2ex]
\end{center}
\caption{\label{fig:trajexamples}\small%
\textbf{Examples of the self-colliding trajectories.}
Diagrams show trajectories with various choices of the impact parameter ${\rho}$, or, equivalently, with a different scattering angle ${\omega}$. The angle $\omega$ also parameterizes the position of the self-collision on the collision circle. Diagrams (a)-(c) represent typical cases ${\omega\in(0,\frac\pi4)}$, ${\omega\in(\frac\gamma2,\frac\pi2)}$, and ${\omega\in(-\frac\pi2,-\frac\gamma2)}$, respectively. The diagram in
the first column depicts the trajectory in the totally covering space, the second column shows the trajectory in the conical space with the wormhole centered on the direction of the incoming trajectory, ${\psi=0}$, the third and the fourth columns correspond to other orientations of the wormhole.
The diagrams (d)-(g) depict special choices of the trajectory, namely those with ${\omega=\frac\gamma2}$, ${\omega=\omega_\fr}$, ${\omega=-\omega_\fr}$, and ${\omega=-\frac\gamma2}$, respectively; in all of them with the wormhole centered on the incoming trajectory.
The arrows indicate passages through the wormhole; however, they do not count the time-shift, since the particle can travel through the time machine in both directions.
We can observe that the structure of the trajectory can change substantially with various choices of the impact parameter and of the incoming direction with respect to the wormhole. For example, the particle can self-collide after passing through the wormhole, both into the past and future, or it can move through the wormhole after the self-collision. }\vspace*{-3ex}
\end{figure*}

It means, that particles which approach the time machine along the same direction ${\ph=0}$ (with various ${\rho\in(-\rho_\mx,\rho_\mx)}$, forming thus congruence with ``plane-wave'' wavefront) self-collide on the common colliding circle and, in the end, leave in the directions which point out from the common point ${S'}$, cf.\ Fig.~\ref{fig:congruence}. In the covering space, the original plane-wave congruence of particles scatters to the circular-wave congruence, but the particles are phase shifted. They fly in radial directions, but the wavefront of the congruence at one moment does not form a circle since the particles start to move in the radial directions in various times. Indeed, the incoming particles do not self-collide at the same time: they reach the collision circle in various times.

Let us mention an interesting feature of the trajectories of the discussed congruence: all inner trajectories (between the self-collisions) have the direction going through one focusing point ${F}$, cf.\ Fig.~\ref{fig:congruence}. For ${\omega\in(-\frac\gamma2,\frac\gamma2)}$ they even pass through this point. The congruence thus focuses in this point, however, the particles do not pass the focusing point at the same moment.

We have obtained a simple picture of the motion in the covering space. It complicates slightly if we return to the conical space by choosing the orientation of the wormhole (i.e., setting mouths at the angles ${-\frac\gamma2+\psi}$ and ${\frac\gamma2+\psi}$ with ${\psi\in(-\frac\gamma2,\frac\gamma2)}$) and making the identification of points with ${\ph}$ differing by an integer multiple of ${\gamma}$. The resulting picture depends on the values of ${\gamma}$, ${\psi}$ and of the impact parameter~${\rho}$; some of the representative trajectories are depicted in Fig.~\ref{fig:trajexamples}.

The typical quantity, which depends on the angular position of the self-collision, is the total time $T$ gained in the wormhole during the whole scattering process. It can be read out in the covering space from the angular coordinate ${\ph_\out = 2\omega+\gamma}$ of the outgoing trajectory:\footnote{%
We have chosen the incoming trajectory with ${\ph_\inc=0}$ and its angular position with respect to the wormhole is given by the parameter ${\psi}$. Alternatively, we could set ${\psi=0}$ and admit ${\ph_\inc\neq0}$.} %
if it belongs to ${n}$-th copy of the wormhole in the covering space, $n\in\textbf{Z}$, the particle gains the time-shift ${T=-n\Delta t}$:
\begin{equation}\label{eq:totts}
   \ph_\out\in\Bigl(-\frac\gamma2+\psi+n\gamma,\frac\gamma2+\psi+n\gamma\Bigr)
   \;\Rightarrow\; T=-n\Delta t\;.
\end{equation}
Taking into account the restrictions on ${\gamma}$, ${\psi}$ and ${\omega}$, the total time-shift can be ${-3\Delta t\leq T\leq\Delta t}$. See Fig.~\ref{fig:trajexamples} for examples.

The total time-shift can be similarly calculated for the collision-free trajectories. In this case the direction of the outgoing trajectory in the covering space is ${\ph_\out=\pi}$ for the trajectories passing the wormhole to the past, and ${\ph_\out=-\pi}$ for the trajectories passing the wormhole to the future. The condition \eqref{eq:totts} gives that the trajectory traveling to the past, ${\rho>0}$, gains the time-shift ${T=-\Delta t}$ for ${\psi\in(-\frac\gamma2,\frac32\gamma-\pi)}$, or it can pass the wormhole twice, ${T=-2\Delta t}$, if ${\psi\in(\frac32\gamma-\pi,\frac\gamma2)}$. Similarly, for ${\rho<0}$, the particle gains the time-shift ${T=\Delta t}$ for ${\psi\in(\pi-\frac32\gamma,\frac\gamma2)}$ and ${T=2\Delta t}$ for ${\psi\in(-\frac\gamma2,\pi-\frac32\gamma)}$.

\section{Conclusion}
\label{sc:concl}

We have analyzed a simple interacting system in the space with a nontrivial causal structure. We have assumed the principle of self-consistency, i.e., we have required the validity of standard local physical laws and we searched for consistent global solutions. In such setting one can formulate the analogue of the so called ``grandfather paradox'' as a question whether all natural initial conditions lead to the global consistent solutions of the local physical laws. Initial conditions which do not lead to such solutions would be paradoxical and the system with nontrivial set of paradoxical initial conditions would be logically inconsistent.

We have shown that the investigated system of finite billiard balls in the non-relativistic space with a time machine realized by a wormhole with a time-shift is not logically inconsistent in this sense. A na\"{\i}ve guess can identify potentially paradoxical initial conditions for which the ball sent through the time machine hits itself and changes inconsistently its own motion. The detailed analysis has shown that even for such initial conditions there exists a global solution satisfying locally all physical laws.

This result was anticipated since it agrees with the previous results obtained for various other systems, in particular, for the very closely related system studied in \cite{billiard}. Our results thus endorse one of the main messages of the study of the system with a nontrivial causal structure: that the presence of time machines does not necessarily imply a drastic reduction of space of the classical solutions of the equations of motion. However, the previous results have been obtained mainly for linear (non-interacting) systems or through a rather complicated and cumbersome analysis of interacting systems. Therefore, the confirmation for the interacting system allowing a detailed explicit analysis is valuable.

We have confirmed also another similarly interesting result discussed in \cite{billiard}. The presence of the nontrivial causal structure in our system has enlarged the space of solution. We have found that for a wide class of initial conditions (namely for ${|\rho|<\rho_\mx}$ with given ${u}$) more than one classical evolutions exist. In addition to the ``standard'' collision-free trajectory the particle can also move along two different self-colliding trajectories, see Fig.~\ref{fig:p.two.solutions}. The evolution of the system thus fails to be deterministic and the classical theory does not have any means to determine which of the different evolutions would be realized.

One can speculate that the quantum theory would be a more complete description. It does not determine exact trajectory of the particle---it just estimates a probability for various possible evolutions (specified up to a quantum uncertainty). In the classical limit only evolutions close to the solutions of the classical laws would have a non-trivial probability. In systems with standard causal structure, there is usually only one such a solution. However, in our system, one can expect that all three classical solutions would have non-trivial probabilities and a semi-classical approximation could give an estimate for these probabilities.

Nevertheless, one has to emphasize that it's not trivial at all to complete this program. The status of quantum theory in the presence of time machines is unclear. The common formulations of the quantum theory are heavily based on the standard notion of time, so a nontrivial causal structure would change the theory substantially.\footnote{%
As an example of possible difficulties let us mention that one would have to take into account the quantum space for degrees of freedom ``hidden inside'' the time machine, i.e., the degrees of freedom which are not encoded in the standard initial conditions. We ignored them on the classical level---they correspond, e.g., to a mysterious particle which appears from the wormhole, hits the particle coming from infinity, and deflects itself back to the time machine in such a way that it reappears from it exactly as the introduced mysterious particle. On quantum level such degrees of freedom cannot be easily ignored or separated.} %
It would be extremely interesting to find a modification of the quantum mechanics for a space with time machines, but it is a hard challenge for further work. The simplicity of our model could give hope that such a challenge is treatable.

However, one could  ask what a key ingredient is for the existence of more solutions of the classical equations of motion. Is it really the presence of the time machine? One could also suspect a peculiar geometrical structure of our conical space. The conical space breaks the uniqueness of the classical solutions even without the time machine. Indeed, for given initial and final points, in the conical space there exist more trajectories which join them. This effect is purely geometrical and does not need a nontrivial causal structure. However, we have identified a different kind of uniqueness: we have found that the specification of both the initial position and the momentum (the complete initial conditions) admits more different evolutions, which is certainly surprising when the local evolution is given by standard differential equations which are generally accepted as deterministic. Non-uniqueness appears here as a strange result of the interplay between a local deterministic evolution and a nontrivial causal structure on a global scale for which the time machine is the key ingredient.

Let us finally mention, that further study \cite{Dolansky:thesis} of our system revealed, that if one takes into account a possibility of multiple self-collisions, the situation gets even more interesting. In this work we have studied only the collision-free trajectories, and the trajectories with one self-collision. However, the incoming and outgoing trajectories from the self-collision can easily self-intersect and for a special choice of parameters even self-collide again. It is possible to show that when the impact parameter ${\rho}$ gets smaller (with fixed ${u}$), there exist more and more multi-self-colliding solutions. For a point-like particle, there exist a finite value ${\rho_{\mx\infty}}$ under which there exist infinite number of possible evolutions for single initial parameters ${\rho}$ and ${u}$, cf.\ \cite{Dolansky:thesis}. We leave further details to the subsequent publication.

Summarizing, the studied system with a wormhole-based time machine does not suffer from the paradoxical initial conditions. On the contrary, it breaks the deterministic character of the theory and offers more solutions for single initial conditions.


\begin{acknowledgments}
P.~K. was supported by Grant No.~GA\v{C}R-202/09/0772 and by Project No.~MSM0021620860. The authors also thank prof. Ji\v{r}\'{\i} Langer for reading the manuscript.
\end{acknowledgments}

\vfill


\end{document}